\documentclass[a4paper,11pt]{article}
\usepackage{jheppub} % for details on the use of the package, please see the JINST-author-manual
\usepackage{lineno}
\usepackage{amssymb}
\usepackage{multirow}
\usepackage{amsmath}
\usepackage{mathrsfs}
\usepackage{hyperref}
\newcommand\tr{\text{Tr}}

\title{\boldmath Some Lower Dimensional Quantum Field Theories Reduced from Chern-Simons Gauge Theories}

\author[a]{Burak Oğuz}
\author[a]{Bayram Tekin}
\affiliation[a]{Department of Physics,
Middle East Technical University,\\
06800 Ankara, Turkey}

\emailAdd{oguz.burak@metu.edu.tr}
\emailAdd{btekin@metu.edu.tr}

\abstract{We study symmetry reductions in the context of Euclidean Chern-Simons gauge theories to obtain lower dimensional field theories. Symmetry reduction in certain gauge theories is a common tool for obtaining explicit soliton solutions. Although pure Chern-Simons theories do not admit solitonic solutions, symmetry reduction still leads to interesting results. We establish relations at the semiclassical regime between pure Chern-Simons theories on $S^3$ and the reduced Quantum Field Theories, based on actions obtained by the symmetry reduction of the Chern-Simons action, spherical symmetry being the prominent one. We also discuss symmetry reductions of Chern-Simons theories on the disk, yielding $BF$-theory in two dimensions, which signals a curious relationship between symmetry reductions and the boundary conformal field theories. Finally, we study the Chern-Simons-Higgs instantons and show that under certain circumstances, the reduced action can formally be viewed as the action of a supersymmetric quantum mechanical model. We discuss the extent to which the reduced actions have a fermionic nature at the level of the partition function.  \\  
}

\begin{document}

\maketitle

\flushbottom

\section{Introduction}
\label{sec:intro}

In gauge theories, one encounters complicated tensorial equations of motion which are hard to solve without some symmetry assumptions. One often uses symmetries to reduce the tensorial equations to scalar equations that are amenable to exact or numerical solutions. Of course, such ans\"atze do not exhaust all solutions since the solutions to the equations of motion need not have the symmetry imposed on the ansatz. But it is useful in any case to obtain some class of solutions that obey some symmetries. Such symmetry reduction method at the level of the action was used by Weyl to obtain the spherically symmetric metric solutions to Einstein's field equations. Weyl's use of symmetry reduction was rather heuristic without proof. Later, it was rigorously proven by Palais \cite{Palais:1979rca} that the symmetric critical points of the action are critical symmetric points of the field equations if the symmetry group is compact. That means that one can find the equations of motion from the full action, apply the symmetric ansatz, and have a reduced equation that can hopefully be solved; on the other hand, one could equivalently insert the ansatz into the full action, integrate out the irrelevant coordinates using the symmetry, obtain a lower-dimensional theory, and solve the equations of the reduced theory. Palais showed that the two paths give the same equations of motion when the symmetry imposed is associated with a compact group such as the rotation groups. \\

Given that the equations of motion of Yang-Mills theory are extremely complicated to solve due to their non-linearity, the trick described above was used to reduce the instanton equations to simpler equations. Witten \cite{Witten:1976ck} proposed an ansatz that depends only on the Euclidean time $t$ and the radial coordinate $r = (x_ix_i)^{1/2}$ to reduce the instanton equations. The ansatz for the $SU(2)$ non-Abelian gauge field on $\mathbb R^4$ is as follows:
\begin{equation} \label{Witten's instanton ansatz, equation 1.1}
\begin{aligned} 
    A_i^a &= \varepsilon_{iak} \frac{x_k}{r^2} (\varphi_2 - 1) + \frac{\delta^{\perp}_{ia}}{r} \varphi_1 + \frac{x_ix_a}{r^2}  A_1, \\
    A_0^a &= \frac{x_a}{r} A_0,
\end{aligned}
\end{equation}
where $\delta^\perp_{ia} = \left( \delta_{ia} - \frac{x_ix_a}{r^2} \right)$, and $A_0,A_1,\varphi_1$, \& $\varphi_2$ depend only on $r$ and $t \equiv x_4$. When one inserts this ansatz into the Yang-Mills action, the reduced action turns out to be the Abelian-Higgs model on the Poincaré half-plane with the metric $g^{ab} = r^2 \delta^{ab}$:
\begin{equation}
    \begin{aligned}
        S_{\text{YM}} = \frac{1}{4} &\int d^4x F_{\mu\nu}^a F_{\mu\nu}^a \\
        &\xrightarrow{} A = 8\pi \int_{-\infty}^{\infty} dt \int_0^{\infty} dr \left[ \frac{1}{2} (D_\mu \varphi_i)^2 + \frac{1}{8} r^2 F_{\mu\nu}^2 + \frac{1}{4} r^{-2} (1 - \varphi_i \varphi_i)^2 \right],
    \end{aligned}
\end{equation}
where $D_\mu \varphi_i = \partial_\mu \varphi_i + \varepsilon_{ij} A_\mu \varphi_j$ and $F_{\mu\nu} = \partial_\mu A_\nu - \partial_\nu A_\mu$ ($i=1,2$ and $\mu,\nu = 0,1)$. The instanton equations of 4d Yang-Mills theory turn out to be the Bogomolny'i equations of vortices of the 2-dimensional Abelian-Higgs model \cite{Witten:1976ck, Manton:2004tk}. Geometrically, the same equations describe minimal surfaces embedded in $\mathbb R^{1,2}$ \cite{Tekin:2000fp, Comtet:1978ue}. \\

Witten's ansatz breaks the symmetry group $SU(2) \times SO(3)$ to $SU(2)$ by mixing the gauge and space coordinates. There is no symmetry in the $t-$direction. For 3-dimensional Euclidean gauge theories, this ansatz can be applied by ignoring the time component in (\ref{Witten's instanton ansatz, equation 1.1}). In three dimensions, besides the Yang-Mills theory, there is another gauge theory, called the Chern-Simons theory. This theory was extensively studied as a topological field theory \cite{Witten:1988hf}, and has important appearances in 3d gauge theories \cite{Dunne:1998qy}, condensed-matter systems \cite{Tong:2016kpv}, and in many areas of mathematical physics. It is also known that Chern-Simons theory for some gauge groups is related to 3d Einstein gravity with or without a cosmological constant in the first order formalism, as shown by Witten \cite{Witten:1988hc}, once the condition on invertibility of the dreibein is relaxed. Due to the many interesting phenomena Chern-Simons theory entails, we will study it within the framework of symmetry reduction. \\

One of our primary goals in studying the Chern-Simons theory in the symmetry reduction context will be to establish connections between the reduced theories and the original ones at the quantum level. Usually, symmetry reductions are useful for obtaining explicit solitonic solutions, but pure Chern-Simons theory has no such solution. However, Chern-Simons theory offers something special: a space of solutions that has a group structure. The space of classical solutions of Chern-Simons theory is given by the space of $G-$flat connections where $G$ is the gauge group. Depending on the quotiening scheme, this group structure may be preserved or lost in the moduli space of flat connections (the moduli space is obtained by modding out the solution space by the gauge redundancy). We will exploit this group structure in the solution space to get an interesting perspective on the symmetry reduction framework for Chern-Simons gauge theories. In particular, as we will discuss in section \ref{sec 3: Symmetry Reduction at the Semiclassical Regime}, one may hand-pick the solution space compatible with a symmetry condition, restrict the partition function of the theory to such configurations, and by changing the gauge group of the theory compatible with this symmetry, one may obtain reduced theories that are semiclassically in agreement to the original Chern-Simons theories. By a semiclassical agreement, we mean that the gauge invariant observables of both theories should match in the semiclassical regime. In this paper, we only deal with the identity observable that is the partition function, and we are not aware of a technology to handle the symmetry reductions of Wilson loop operators, so we leave the investigation of that as an open problem. \\

Below, we summarize the sections of the paper:
\begin{itemize}
    \item Section \ref{sec 2: Euclidean SU(2) CS Theory on S1 * S2}: Our starting point is the pure $SU(2)$ Chern-Simons theory in the Euclidean formalism, on a manifold with the topology of $S^1\times S^2$. We employ a 3-dimensional version of spherically symmetric ansatz on the gauge field and show that the reduced action can be identified with a fermionic action coupled to a $U(1)$ gauge field with a 1d Chern-Simons kinetic term. A general version of this model was studied in \cite{Dunne:1996yb}. We argue that at the level of classical equations of motion, one can map the spherically symmetric solutions of the 3d Chern-Simons theory to the classical solutions of the model studied in \cite{Dunne:1996yb}. In subsection \ref{subsec 2.1: S1 * S2 reduction}, we discuss how symmetry reduction works for $S^1 \times S^2$, and evaluate the path integral of the reduced theory, which unusually has a complex bosonic field with a fermionic kinetic term. We show that the partition function gives 1, in agreement with the partition function of Chern-Simons theory on $S^1\times S^2$. In subsection \ref{subsec 2.2: 1d theory path integral as a susy qm index}, we obtain a fermionic theory from the reduced action at the level of the path integral, and we reinterpret the reduced theory as a theory with a boson and a fermion both coupled to the $U(1)$ gauge field with the same charge. The fermionic field has a periodic boundary condition, hence we interpret the partition function as the Witten index of a supersymmetric quantum mechanics with a trivial superpotential. Computationally, this is not interesting; but conceptually it is intriguing as it relates the Chern-Simons partition function with a supersymmetric index. 
    \item Section \ref{sec 3: Symmetry Reduction at the Semiclassical Regime}: We discuss the symmetry reduction of Chern-Simons theories on $S^3$. For $SU(2)$, we get the same reduced action as in the $S^1 \times S^2$ case, but because we are reducing from a different topology, we obtain different boundary conditions on the reduced fields. We study the partition function of the reduced theory and show that it can be recast as a theory with a complex boson and a fermion. In subsection \ref{subsec 3.1: spherical symmetry reduction of SU(2) CS on S^3} we focus on the semiclassical evaluation of the $SU(2)$ Chern-Simons theory on $S^3$. We show that in the semiclassical limit, the partition function of the Chern-Simons theory agrees with that of the reduced theory. We extend this result to argue that the quantum observables of the two must agree as well. This implies an agreement at the semiclassical level between a 3d theory and a theory that can effectively be described by a 1d theory. To what do the quantum observables of the Chern-Simons theory, the Wilson loops, correspond in the reduced theory is an outstanding problem. In subsection \ref{subsec 3.2: SYM reduction of G CS on S^3}, we generalize our arguments for a Chern-Simons theory with a general gauge group $G$ on $S^3$. We also allow for generalizations in the symmetry reduction scheme from spherical symmetry to a general symmetry reduction scheme associated with a compact symmetry group, denoted $\mathcal{S}ym$. We argue that due to the group structure of the space of critical points of the Chern-Simons action, one can hand-pick certain regions in the solution space and restrict the theory to that region compatible with a $\mathcal{S}ym$-reduction, and if the gauge group is changed to be compatible with the $\mathcal{S}ym$-reduction, one can generate families of reduced theories that are semiclassically in agreement with the original theory.
    \item Section \ref{sec 4: SL(2,C) Chern Simons or dS_3 gravity}: We extend our discussion to Euclidean Chern-Simons gauge theory with the gauge group $SL(2,\mathbb C)$. This choice is particularly interesting due to its relation to de Sitter gravity in 3 dimensions \cite{Witten:1988hc, Verlinde:2024zrh}. Our results from $SU(2)$ smoothly extend in a foreseeable manner. The reduced action under a spherical symmetric reduction gives two copies of the reduced action obtained in $SU(2)$, so everything we did for $SU(2)$ appears here with two independent sectors. In subsection \ref{subsec 4.1: Semiclassical Limit of SL(2,C) Chern-Simons Gauge Gheory}, we discuss the semiclassical evaluation of the $SL(2,\mathbb C)$ Chern-Simons theory. This section is an example of the generalization constructed in the subsection \ref{subsec 3.2: SYM reduction of G CS on S^3} with $G = SL(2,\mathbb C)$, and $\mathcal{S} ym$ being a spherical symmetry. In \ref{subsec 4.2: possible applications}, we discuss possible applications and relations of this reduction scheme with double-scaled Sachdev-Ye-Kitaev models (DSSYK), in relation to the recent literature \cite{Verlinde:2024zrh, Verlinde:2024znh, Narovlansky:2023lfz, Susskind:2022bia}.
    \item Section \ref{sec 5: Disk Reduction of Chern-Simons Theories on X = R * D2}: In this section, we digress from spherical symmetry reduction to what we refer to as a disk reduction scheme. We consider the Chern-Simons theories defined over $X =\mathbb R \times D_2$ with $D_2$ being a disk and write down symmetry ans\"atze on the gauge fields that are independent of the angular coordinate on the disk. This disk reduction scheme yields $BF$-theories in 2-dimensional spacetimes. For the $U(1)$ Chern-Simons theory on $X$, we get a $U(1)$ $BF$-theory on $I_R \times \mathbb R$ or $S^1_R \times \mathbb R$ where $I_R = (0,R)$ is the finite interval and $S^1_R$ is a circle of radius $R$. For the $SU(2)$ Chern-Simons theory on $X$, making a disk reduction gives a $U(1)\times U(1)$ $BF$-theory on the same two-dimensional spacetimes. These disk reductions are intriguing as they give theories that are related to the boundary conformal field theories (CFTs) appearing in the $U(1)$ Chern-Simons theory. In any case, based on the results in this section, there is good reason to believe that these disk reductions may shed new light on the interplay between the Chern-Simons gauge theories in the bulk and the boundary CFTs \cite{Kapustin:2010if, Witten:1988hf, Elitzur-Shmuel-Moore-Schwimmer-Seiberg:1989nr, Tong:2016kpv, Dijkgraaf:1989pz}. 
    \item Section \ref{sec 6: SU(2) Chern-Simons Higgs and their monopoles}: Having studied pure Chern-Simons theories, we couple the Higgs field to the gauge sector and study symmetry reduction for these theories. For Euclidean Chern-Simons-Higgs theories, it is known that there are instanton-type solutions, and this model has been studied in the symmetry reduction scheme in the past \cite{Edelstein:1994dy}. We study this based on an expectation that the reduced action under the symmetry ansatz has a supersymmetry-like structure. In subsection \ref{subsec 6.1: CS-Higgs on M = R * S2}, upon exploiting the conformal equivalence $\mathbb R^3 -\{0 \} \sim \mathbb R^+ \times S^2$, we study the Chern-Simons Higgs theory defined on the Euclidean spacetime $\mathcal M = \mathbb R^+ \times S^2$ (where $\mathbb R^+ = (0,\infty)$). Imposing the fields to be independent of $S^2$, we obtain a reduced action on $\mathbb R^+$ that looks like a supersymmetric quantum mechanical model. It is not precisely a supersymmetric quantum mechanics because the reduced action consists of a real scalar field with an ordinary kinetic term and a complex scalar field with a fermionic kinetic term. This is unusual, but at the level of classical equations of motion, one can view this as the equations of supersymmetric quantum mechanics. We conclude that the spherically symmetric instanton equations of the Euclidean Chern-Simons-Higgs theory are formally the same as the classical solutions of supersymmetric quantum mechanics. In subsection \ref{subsec 6.2: A new perspective on the reduced CSH theory}, we try to relate the reduced theory with a fermionic theory at the level of the path integral. We succeed in obtaining an action that contains a real scalar field, a complex scalar field, and a complex Grassmannian field. The action has almost supersymmetry, but the kinetic term of the complex scalar field contains the real scalar field, which breaks the supersymmetry that is present in the rest of the action. 
    \item Section \ref{sec 7: SL(2,C) Chern-Simons Higgs and their monopoles}: We extend our results from the study of the $SU(2)$ Chern-Simons Higgs theory to $SL(2,\mathbb C)$. The transition of the results is again smooth and foreseeable. We get two copies of the reduced action we obtained in section \ref{sec 6: SU(2) Chern-Simons Higgs and their monopoles}.
    \item Section \ref{sec 8: Semiclassical limit of Quantum Chern-Simons Higgs theory}: Having studied the classical Chern-Simons Higgs theory and its instantons, we discuss the quantum theory. Computations are more involved now due to the interactions between the fields. We demonstrate the challenges by studying a 0-dimensional toy model and argue that the semiclassical evaluation of the quantum Chern-Simons Higgs theory is much more complicated than that of the pure Chern-Simons theory. Nevertheless, we sketch a semiclassical argument discussing how much the dimensionally reduced action is relevant for the quantum Chern-Simons-Higgs theory.
    \item We delegate some of the more involved computations to the appendices. 
\end{itemize}

\section{Symmetry Reduction of the Euclidean \texorpdfstring{$SU(2)$}{SU(2)} Chern-Simons Theory} 
\label{sec 2: Euclidean SU(2) CS Theory on S1 * S2}
The Euclidean Chern Simons theory on a 3-manifold $\mathcal M$ is defined by the action
\begin{equation} \label{Action of the 3d Euclidean Chern-Simons theory I}
    I = -\frac{i \kappa}{4\pi} \int_{\mathcal M} CS(A),
\end{equation}
with 
\begin{equation} \label{Chern-Simons form}
    CS(A) = \tr \left( A\wedge dA + \frac{2}{3} A \wedge A \wedge A \right),
\end{equation}
and $A \in \Omega^1(\mathcal M, \mathfrak{su}(2))$ is the $\mathfrak{su}(2)$-valued connection of the Principal $SU(2)$ bundle over $\mathcal M$. We choose the quadratic bilinear form in the Lie algebra $\mathfrak{su}(2)$ such that $\tr(t^at^b) = -\frac{1}{2} \delta^{ab}$. In components, the action reads
\begin{equation}
    I = \frac{i\kappa}{8\pi} \int_{\mathcal M} d^3x \varepsilon^{ijk} \left( A_i^a\partial_j A_k^a + \frac{1}{3} \varepsilon^{abc} A_i^a A_j^b A_k^c \right).
\end{equation}
Under a gauge transformation, the change in the action is 
\begin{equation} \label{variation of the Chern-Simons action}
    \delta I = -2\pi i \kappa n + \frac{i\kappa}{4\pi} \int_{\partial \mathcal M} \tr \Big( A \wedge dg^{-1} g \Big),
\end{equation}
with $n$ being the winding number
\begin{equation}
    n = -\frac{1}{24\pi^2} \int \tr \Big( gdg^{-1} \wedge gdg^{-1} \wedge gdg^{-1} \Big) \in \mathbb Z.
\end{equation}
The second term in the right-hand side of (\ref{variation of the Chern-Simons action}) is a boundary term, which vanishes for a manifold without a boundary. In this section, we take $\mathcal M = S^1\times S^2$. So the gauge invariance of the quantum theory yields the condition 
$\kappa \in \mathbb Z$ as is well-known. \\

Classical field equations
\begin{equation}
    F = 0,
\end{equation}
are solved by flat connections $A = gdg^{-1}$. For the component fields $A_i^a$, we employ the spherically symmetric ansatz that is a 3-dimensional version of the multi-instanton ansatz of \cite{Witten:1976ck}:
\begin{equation} \label{Spherically symmetric ansatz on A}
    A = A_i^a t^a dx^i = t^a \Bigg[ \Big(\varphi_2(r) - 1 \Big) \varepsilon_{iak} \frac{x_k}{r^2} + \varphi_1(r) \frac{\delta^{\perp}_{ia}}{r} +  A(r) \frac{x_ix_a}{r^2} \Bigg] dx^i,
\end{equation}
where $\delta^{\perp}_{ia} \equiv \delta_{ia} - \frac{x_ix_a}{r^2}$, and $r= (x_ix_i)^{1/2}$ ($i=1,2,3$). With this ansatz, one gets (see Appendix for explicit computations):
\begin{equation}
    \frac{1}{2} \varepsilon^{ijk} F_{jk}^a = -\left( \frac{\partial_1\varphi_1  - A \varphi_2}{r^2} \right) \varepsilon_{iak} x_k + \left( 
    \frac{\partial_1\varphi_2 + A \varphi_1}{r} \right) \delta^{\perp}_{ia} - \left( \frac{1 - \varphi_1^2 - \varphi_2^2}{r^4} \right) x_ix_a.
\end{equation}
For the Chern-Simons action, we can integrate over the angles and get a reduced 1-dimensional action (a prime denotes $\partial_r$ and we denote the reduced action by $S_{\text{CS}}$)
\begin{equation} \label{Chern Simons 1d action}
\begin{aligned}
    S_{\text{CS}} = i\kappa \int dr \Big( \varphi_1 \varphi'_2 - \varphi'_1 (\varphi_2-1) + A (\varphi_2^2 + \varphi_1^2 - 1) \Big).
\end{aligned}
\end{equation}
And we drop the boundary term $\int dr \varphi_1'$. This action defines a theory in a 1-dimensional spacetime in the Euclidean formalism. Either from varying this action with respect to $\varphi_1,\varphi_2, A$ or by using the form of $F_{ij}^a$ under the reduction, we have the following field equations
\begin{equation}
\begin{aligned}
    0 &= \partial_1 \varphi_1 - A \varphi_2 , \\
    0 &= \partial_1 \varphi_2 + A \varphi_1 , \\
    0 &= \varphi_1^2 + \varphi_2^2 -1 .
\end{aligned}
\end{equation}
We combine $\varphi_1$ and $\varphi_2$ into a complex scalar field $Y$ via:
\begin{equation}
\begin{aligned}
    Y \equiv \varphi_1 + i \varphi_2 \quad ; \quad 
    \overline Y \equiv \varphi_1 - i \varphi_2,
\end{aligned}
\end{equation}
with this change of variables and an integration by parts,  (\ref{Chern Simons 1d action}) becomes
\begin{equation} \label{Reduced action in terms of Y}
    S_{\text{CS}} = \kappa \int dr \Big( \overline Y (\partial_r + iA ) Y - i A \Big),
\end{equation}
which resembles the action of a fermion living in 1-dimension, coupled to a gauge field $A$ of which the kinetic term is a 1-dimensional Chern-Simons term. Assuming $\kappa >0$, and rescaling $Y$ with $\sqrt \kappa$, one can write this as
\begin{equation} \label{CS reduced in terms of complex scalar field}
    S_{\text{CS}} = \int dr \Big( \overline Y (\partial_r + i A) Y - i\kappa A \Big).
\end{equation}
For the path integral based on the action (\ref{CS reduced in terms of complex scalar field}) to be invariant under large $U(1)$ gauge transformations, $\kappa$ must be an integer, the same condition coming from the 3d theory. The fields change under $U(1)$ transformations as
\footnote{Under the reduction, there is a surviving $U(1)$ subgroup of $SU(2)$ for which the elements can be written $U = \exp \left( -i/2 f(r)  \hat x\cdot \sigma  \right)$. It is under these transformations the above action is invariant up to $2\pi n$, $n$ being the winding number.}
\begin{equation}
\begin{aligned}
    Y &\xrightarrow{} e^{-i \Lambda(r)} Y, \\
    A & \xrightarrow{} A + \partial_r \Lambda = e^{-i\Lambda} (A - i  \partial_r ) e^{i\Lambda}.
\end{aligned}
\end{equation}

There is a striking resemblance of the kinetic term of $Y$ in (\ref{CS reduced in terms of complex scalar field}) to the kinetic term of a fermion $\psi$ living in 1d. If we identify $\tau = r$ and associate to the complex scalar field $Y(r)$ a Grassmannian valued field $\psi(\tau)$, we get the following Euclidean action
\begin{equation} 
    S_{\text{CS}} = \int d\tau \Big( \overline \psi (\partial_\tau + iA) \psi - i \kappa A \Big).
\end{equation}
Hence, the classical solutions of the Chern-Simons theory that possess spherical symmetry can formally be viewed as the classical solutions of a fermionic theory, in the sense that the solutions can be mapped to each other. This observation is valid only classically, what about the quantum level? Do we have a similar situation there? In subsection \ref{subsec 2.2: 1d theory path integral as a susy qm index}, we answer this question, where we show that the partition function based on the action (\ref{Reduced action in terms of Y}) can be recast as that of a theory with a boson and a fermion, both coupled to the $U(1)$ field with the same charge.

\subsection{Reduction from \texorpdfstring{$S^1\times S^2$}{S1 x S2}} \label{subsec 2.1: S1 * S2 reduction}
In applying the symmetry reduction from the spacetime $S^1\times S^2$, we are using the conformal equivalence
\begin{equation}
    \mathbb R^3 - \{ 0 \} \sim \mathbb R^+ \times S^2,
\end{equation}
where $\mathbb R^+ = (0,\infty)$. To see this, we look at the metric of $\mathbb R^3$ with $r>0$:
\begin{equation} \label{conformal equivalence of R^3 metric with R+ * S^2 metric}
    ds^2 = dr^2 + r^2 d\Omega^2 = r^2 \left( \frac{dr^2}{r^2} + d\Omega^2 \right),
\end{equation}
which establishes the conformal equivalence above. This is analogous to what happens in the multi-instanton ansatz \cite{Witten:1976ck}, where the conformal equivalence $\mathbb R^4 - \mathbb R \sim \mathbb H^2 \times S^2$ is used \cite{Manton:2004tk}, with $\mathbb H^2$ being the Poincaré upper half-plane. Getting back to our case, if we compactify the radial coordinate, we obtain $S^1 \times S^2$ with the circle having radius $\beta$. After the symmetry reduction of the Chern-Simons term, we consider the following path integral with the periodic boundary conditions on the fields $Y(\beta) = Y(0)$
\begin{equation}
    Z(S^1 \times S^2 \xrightarrow{} S^1) = \int \mathcal D A\mathcal D Y \mathcal D \overline Y \bigg|_{Y(\beta)= Y(0)} \exp \left( -\kappa \int dr \Big( \overline Y D Y - iA \Big) \right),
\end{equation}
where $DY = \partial_r Y + iA Y$. We rescale $X = \sqrt\kappa Y$ to take $\kappa$ out of the kinetic term. The $Y$ field can be expanded in eigenfunctions of $D$, denoted by $\chi_n$, where $n\in \mathbb Z$ due to the boundary conditions. Thus, the measure changes as $\mathcal D X \mathcal D \overline X = \kappa^{2N+1} \mathcal D Y \mathcal D \overline Y$ with $N$ an integer that we will send to $\infty$. The naive limit does not make sense, so we write it as $N = \sum_{m=1}^{N} 1$ and the discrete limit $\lim_{N\to \infty} N$ corresponds to $\zeta(0)= -1/2$, where $\zeta(s)$ is the Riemann zeta function. Hence, $\mathcal D X \mathcal D \overline X = \mathcal D Y \mathcal D \overline Y$ and we have
\begin{equation}
    Z(S^1 \times S^2 \xrightarrow{} S^1) = \int \mathcal D A\mathcal D X \mathcal D \overline X \bigg|_{X(\beta)= X(0)} \exp \left( - \int dr \Big( \overline X D X - i \kappa A \Big) \right).
\end{equation}
The overall path integral seems to depend on $\kappa$, but this is a red-herring: The saddle points of the above action can be given by $X = X_0 e^{-if(r)}$, $A= f'(r)$ for some function $f$, and the periodicity on $X$ imposes $f(\beta) - f(0) = 2\pi n$, and so $i\kappa \int_0^\beta dr A(r) = i\kappa 2\pi n$. Since $\kappa$ is an integer, we have $e^{2\pi i n \kappa} = 1$, independent of $\kappa$ as promised. The fluctuations around classical solutions bring  $1/\det D$, which is also independent of $\kappa$. Since the theory is quadratic, the saddle point evaluation gives the exact value for $Z$. Having established that $Z$ is independent of $\kappa$, we have $Z(S^1 \times S^2 \xrightarrow{} S^1) = 1$ with some appropriate normalization. \\

This is in agreement with the result for the 3d Chern-Simons theory with the topology of $S^1\times S^2$ \cite{Witten:1988hf}, which has $Z=1$ for all $\kappa$ (and for all gauge groups $G$). Our reduction scheme is compatible with this, suggesting that the quantum theory of the reduced action captures some aspects of the $3d$ Chern-Simons theory defined on  $S^1\times S^2$. In the next section, we will argue that the semiclassical (large $\kappa$) limit of the 3d Chern-Simons theory on $S^3$ agrees with the large $\kappa$ limit of the reduced action on the saddle point.

\subsection{A New Perspective on the 1d Theory} \label{subsec 2.2: 1d theory path integral as a susy qm index}
The 1d theory has the following path integral after the integrations over $Y$ are performed
\begin{equation}
    Z = \frac{1}{Z_0} \int \mathcal D A \frac{1}{\det \kappa D} e^{i\kappa \int dr A(r)},
\end{equation}
where $D = \partial_r + i A$. The appearance of the determinant is not familiar. This is because we have a Dirac operator acting on a \emph{scalar} field, which unusually gives the determinant of the Dirac operator in the denominator. Although this appears unnatural, we can recast this to look like a familiar path integral by rewriting the determinant as
\begin{equation}
    Z = \frac{1}{Z_0} \int \mathcal D A \frac{\det(-\kappa D)}{\det( - \kappa^2 D D)} e^{i\kappa \int dr A(r)}.
\end{equation}
We can introduce a complex scalar field $X$ and a complex Grassmannian field $\psi$ to expand the determinants in terms of functional integrals:
\begin{equation}
\begin{aligned}
    Z &= \frac{1}{Z_0} \int \mathcal D A \mathcal D X \mathcal D \overline X \bigg|_{X(\beta) = X(0)} \mathcal D \psi \mathcal D \overline \psi \bigg|_{\psi(\beta) = \psi(0)} e^{-S_E},\\
    S_E &= \int dr \Big( \kappa^2 \overline{DX} D X - \kappa \overline \psi D \psi - i\kappa A \Big).
\end{aligned}
\end{equation}
This theory has a bosonic and a fermionic field, both having the same charge under the $U(1)$ coupling, but the sign of the fermionic term is wrong. To cure this, we can change $\kappa$ to $-\kappa$ as long as it is an integer, which remedies the kinetic term of $\psi$, but the sign of the Chern-Simons kinetic term has now changed. The action becomes
\begin{equation}
    S_E = \int dr \Big( \kappa^2 \overline{DX} D X + \kappa \overline \psi D \psi + i\kappa A \Big).
\end{equation}
We could also redefine the $U(1)$ field as $A \xrightarrow{} -A$, which flips the charge of the fields under $U(1)$, but this is not crucial so we leave it as it is. If we write $X = x_1 + ix_2$, we get the following Lagrangian
\begin{equation}
    L = \kappa^2 ( (Dx_1)^2 + (Dx_2)^2 ) + \kappa \overline \psi D \psi + i \kappa A.
\end{equation}
It is possible to interpret this as a supersymmetric theory 
\begin{equation}
\begin{aligned}
    L &= \kappa^2 ( (Dx_1)^2 + (Dx_2)^2 ) + \kappa \overline \psi D \psi + i \kappa A \\
    & \hspace{2cm} + \left( \frac{dh_1}{dx_1} \right)^2 + \left( \frac{dh_2}{dx_2} \right)^2 + \frac{d^2 h_1}{dx_1^2}\ \overline\psi \psi + \frac{d^2 h_2}{dx_2^2}\ \overline\psi \psi,
\end{aligned}
\end{equation}
 with two independent superpotentials $h_1,h_2$, both coupled to the same fermionic field $\psi$, and both are trivial in the sense that there is no interaction between the boson and the fermion. This means that both superpotentials are trivial $x_i$: $h_i =$ constant. The partition function above is then in some sense a Witten index for a supersymmetric theory with a trivial superpotential, because of the periodic boundary conditions on $\psi$ \cite{Hori:2003ic}. Although this perspective is not very useful computationally in the present context, it is conceptually intriguing as it relates to the partition function of Chern-Simons theory, which is supposed to be a topological invariant of the $3$-manifold, to a supersymmetric index. 

\section{Symmetry Reduction at the Semiclassical Regime} \label{sec 3: Symmetry Reduction at the Semiclassical Regime}
In this section we investigate the symmetry reduction of the Chern-Simons Theory on $S^3$. Because the Chern-Simons action is metric independent, the reduced action for $S^1\times S^2$ and $S^3$ will be the same, given by (\ref{Reduced action in terms of Y}). However, the change of topology will be encoded in the boundary conditions. For reduction from $S^3$, we do a one-point compactification of $\mathbb R^3$ and impose the gauge field $A_i^a$ to be a flat connection at the point at infinity: $A_i^a(\infty) t^a dx^i = UdU^{-1}$. Because we are making a spherically symmetric reduction, we choose $U$ to be a radially symmetric gauge element $U = \exp (f(r) \frac{x^a}{r} t^a)$ with $f$ a function depending only on $r = (x_i x_i)^{1/2}$ $(i=1,2,3)$. For such $U$, one has
\begin{equation}
    UdU^{-1} = t^a \left[ \varepsilon_{iak} \frac{x_k}{r^2} (\cos f - 1) + \frac{\delta^\perp_{ia}}{r} \sin f + \frac{x_ix_a}{r^2} f' \right] dx^i.
\end{equation}
Comparing with (\ref{Spherically symmetric ansatz on A}), we read the boundary conditions at infinity on the ansatz functions as $\varphi_1(\infty) = \sin f(\infty)$, $\varphi_2(\infty) = \cos f(\infty)$, $A(\infty) = f'(\infty)$, and in terms of the $Y$ field $Y(\infty) = e^{i\pi} e^{-if(\infty)}$. We also need boundary conditions at $r=0$ because we did not take that out of the manifold. We choose the boundary conditions at $r=0$ such that $A_i^a$ is regular there, which leads to $\varphi_1(0) = 0$, $\varphi_2(0) = 1$, and hence $Y(0) = e^{i\pi}$. This boundary condition can be written as $Y(\infty) = e^{-if(\infty)} Y(0)$, and since $f(\infty)$ is related to the integral of the gauge field $\int_0^\infty dr A(r)$ in the saddle point, this is like a twisted boundary condition. With these boundary conditions, we have a reduced theory on the closed interval $I_\infty = [0,\infty]$, with the boundary $\partial I_\infty = \{ 0 \} \cup \{ \infty \}$. Since we have a boundary, the boundary term $i\kappa \int dr \partial_r \varphi_1$ that we dropped for $S^1\times S^2$ becomes nonzero for $S^3$. From the boundary conditions on $\varphi_1$, we see that the boundary term evaluates to $i\kappa \int dr \partial_r \varphi_1 = i\kappa \sin f(\infty)$. So, the partition function of the reduced theory reads
\begin{equation} \label{Path integral of the S3 reduced theory}
    Z(S^3 \xrightarrow{} I_\infty ) = \frac{1}{Z_0} \int \mathcal DA \mathcal D Y \mathcal D \overline Y \bigg|_{Y(\infty) = e^{-if(\infty)} Y(0)} e^{-\kappa \int dr( \overline Y D Y -i A) \ -i\kappa \sin f(\infty) }.
\end{equation}
To carry out the saddle-point evaluation, we observe that they are given by $Y = e^{i\pi} e^{-if(r)}$, $A= f'(r)$ so the on-shell action reads $i\kappa \big (f(\infty)- \sin f(\infty) \big)$ (where we use $f(0) =0$), and the fluctuations around the classical solutions bring down $1/\det (\kappa D_{f'})$, where $D_{f'}= \partial_r + if'$ is the second derivative of the action evaluated on the classical solution. Thus, we have
\begin{equation}
    Z(S^3 \xrightarrow{} I_\infty) = \frac{1}{Z_0} \int \mathcal D f e^{i\kappa \big( f(\infty) - \sin f(\infty) \big)} \frac{1}{\det \kappa D_{f'}}.
\end{equation}
We first observe that \cite{Dunne:2000if}
\begin{equation}
    N = -\frac{1}{24\pi^2 } \int_{S^3} \tr \left( Ud U^{-1} \wedge Ud U^{-1} \wedge Ud U^{-1} \right) =  \frac{1}{2\pi} \Big[ f(\infty) - \sin f(\infty) \Big],
\end{equation}
hence the exponential of the on-shell action gives $e^{2\pi i N \kappa} = 1$. To evaluate the determinant, we first solve the eigenvalue equation $\kappa D_{f'} \chi_n = i\lambda_n \chi_n$ ($D$ is an anti-hermitian operator, so its eigenvalues are purely imaginary hence $\lambda_n$ will be real with this convention), with the boundary condition $\chi_n(\infty) = e^{-if(\infty)} \chi_n(0)$. We have
\begin{equation}
    \lambda_n = \frac{\kappa}{R} 2\pi n, \quad n \in \mathbb Z - \{0\}
\end{equation}
where we take $R$ to be the length of the interval $I_\infty$, for which we will take the limit $R\xrightarrow{} \infty$ at the end. Note that $n=0$ is a zero mode, so we need to treat it separately. The zero modes are normalized with a factor of $\frac{1}{\sqrt R}$ so that the inner product of $\overline\chi_0$ and $\chi_0$ is properly normalized. Then, the contribution of the zero mode to the determinant is $\int \mathcal D\overline \chi_0 \mathcal D \chi_0 = \frac{1}{R}$. \\

To evaluate the contributions of the nonzero modes, we define $\zeta_D(s) = \sum_{n\in \mathbb Z -\{0\} } \frac{1}{(i\lambda_n)^s} = \sum_{n = -\infty}^{-1} \frac{1}{(i\lambda_n)^s} + \sum_{n=1}^{\infty} \frac{1}{(i\lambda_n)^s}$, and in the first sum we change $n$ to $-n$ and write 
\begin{equation}
    \zeta_D(s) = \left( \frac{2\pi i\kappa}{R} \right)^{-s} \Big( (-1)^{s}+1 \Big) \zeta(s),
\end{equation}
where $\zeta(s) = \sum_{n=1}^{\infty} \frac{1}{n^s}$ is the Riemann's zeta function. We can thus write the determinant in terms of $\exp( -\zeta'_D(0))$. Using the analytic continuation of Riemann zeta function to $s=0$, $\zeta(0) = -\frac{1}{2}$, $\zeta'(0) = -\frac{1}{2} \ln 2\pi$, we get
\begin{equation}
    \frac{1}{\det (\kappa D_{f'} )} = \frac{1}{R} \left( \prod_{n\neq 0} i\lambda_n \right)^{-1} = \frac{1}{\kappa},
\end{equation}
so we have
\begin{equation}
    Z(S^3 \xrightarrow{} I_\infty) = \frac{1}{\kappa}.
\end{equation}
Just as in section \ref{subsec 2.2: 1d theory path integral as a susy qm index}, we can recast this path integral in a more familiar way. Right now, we have a fermionic kinetic term on a scalar field. Integrating over $Y$, we get $\det(\kappa D)^{-1}$, and we can write this as $\det(-\kappa D)/ \det(-\kappa^2 D D)$ which can be written, with the introduction of new fields, as
\begin{equation} \label{Zspherical as a bosonic + fermionic U(1) theory in 1d}
    Z(S^3 \to I_\infty) = \frac{1}{\text{vol} (\mathcal G_0^r) } \int \mathcal D A \mathcal D X \mathcal D \overline X \bigg|_{X(\infty) = e^{-if(\infty)} X(0)} \mathcal D \psi \mathcal D \overline \psi \bigg|_{\psi(\infty) = e^{-if(\infty)} \psi(0)} e^{-S_E}.
\end{equation}
After flipping the sign of $\kappa$ so that the fermionic kinetic term has the correct sign, $S_E$ reads
\begin{equation}
    S_E = \int dr \Big( \kappa^2 \overline{DX} D X + \kappa \overline \psi D \psi - i\kappa A \Big) + i\kappa \sin f(\infty).
\end{equation}
As we did in section \ref{subsec 2.2: 1d theory path integral as a susy qm index}, we can interpret the path integral in (\ref{Zspherical as a bosonic + fermionic U(1) theory in 1d}) as the index of a supersymmetric quantum mechanics with trivial superpotentials, but now with twisted boundary conditions. This may lead to a conceptually profound relation between the path integral of the Chern-Simons theory and a supersymmetry index and should be explored. \\

\subsection{Spherical Symmetry Reduction of Quantum Chern-Simons Theory on \texorpdfstring{$S^3$}{S**3} with Gauge Group \texorpdfstring{$SU(2)$}{SU(2)} } \label{subsec 3.1: spherical symmetry reduction of SU(2) CS on S^3}

We will try to establish a relation between the path integral of the Chern-Simons theory on $S^3$ and the quantum theory defined by the partition function in (\ref{Path integral of the S3 reduced theory}). Note that the classical solutions of the reduced theory solve the equations of the 3-dimensional Chern-Simons theory, however, the converse is not true. The solutions of the reduced theory do not exhaust all of the solutions of the 3-dimensional theory, because not all solutions have to be spherically symmetric in 3d. Since some space of the classical solutions of the two agrees, one would expect an approximate relation at the semiclassical regime, which corresponds to the large $\kappa$ limit of both (\ref{Action of the 3d Euclidean Chern-Simons theory I}) and (\ref{Reduced action in terms of Y}). \\

It turns out, as we will argue below, that both theories agree in the semiclassical regime. This agreement holds not only for the partition functions but also for the quantum observables. For a starting point, we will study a toy model. Suppose we are interested in the following integral defining a 0-dimensional QFT:
\begin{equation}
    Z = \int dx \ e^{-f(x)}.
\end{equation}
We denote the space of solutions to $\frac{df}{dx} = 0$ by $\mathscr M$. For $x_i \in \mathscr M$, one expands $f(x)$ to second order in the parameter $\delta x = x-x_i$
\begin{equation}
    f(x) = f(x_i) + \frac{1}{2} f''(x_i) \delta x^2,
\end{equation}
then one can approximate the partition function by summing over all $x_i \in \mathscr M$ 
\begin{equation} 
    Z \approx \sum_{x_i \in \mathscr M} e^{-f(x_i)} \int_{B(x_i)} d[\delta x] e^{-\frac{1}{2} f''(x_i) \delta x^2},
\end{equation}
with $B(x_i)$ being some neighborhood of the classical point $x_i$. If $B(x_i)$ contains the whole of the real line, then one can perform the Gaussian integral, but the expansion to the second order only holds for small enough $\delta x$. One can avoid this pitfall by considering the limit in which a parameter appearing in the classical action $f(x)$ to be very large so that the Gaussian is sharply peaked around the minima and quickly decays hence the integral around $B(x_i)$ is reasonably close to the integral over the entire real line. In that limit, the semiclassical limit, one has
\begin{equation}
    Z \approx \sum_{x_i \in \mathscr M} e^{-f(x_i)} \Big( f''(x_i) \Big)^{-1/2}.
\end{equation}
Now, we will apply this strategy verbatim in the quantum Chern-Simons theory defined over the $S^3$ topology. The Euclidean $SU(2)$ Quantum Chern-Simons Theory over $\mathcal M = S^3$ is defined by the path integral
\begin{equation}
    Z(S^3) = \int \mathscr D A \exp \Big( -I[A] \Big) = \int \mathscr D A \exp \left( \frac{i\kappa}{4\pi} \int_{\mathcal M} CS(A) \right),
\end{equation}
with $CS(A)$ given as in (\ref{Chern-Simons form}), and $A \in \Omega^1(\mathcal M, \mathfrak{su}(2))$. As discussed in the beginning, the quantum theory is gauge invariant provided $\kappa \in \mathbb{Z}$. To carry out the semiclassical approximation, we first need to determine the space of stationary points of $I$. The stationary points are given by the flatness condition:
\begin{equation}
    F = 0,
\end{equation}
which are solved by pure gauges $A = gd g^{-1}$. Hence, the space of critical points of the Chern-Simons action is given by
\begin{equation}
    \mathscr M_{\text{CS}} = \left\{ \mathcal A(x) = g(x) d g^{-1}(x) \ \bigg| \ g \in SU(2) \right\},
\end{equation}
which is isomorphic to the space of gauge transformations $\mathcal G$. The isomorphy is established by the pure gauge connections: $g(x)  \mapsto \mathcal A(x) = g(x) d g^{-1}(x)$. 
To obtain the moduli space of flat connections, we quotient out the gauge redundancy from $\mathscr M_{\text{CS}}$. There are two quotiening schemes, which lead to different moduli spaces. The first quotiening scheme is modding out the space of all gauge transformations $\mathcal G = \oplus_{n=-\infty}^{\infty} \mathcal G_n$, where $\mathcal G_n$ is the space of gauge transformations of winding number $n$, in which case there is only one flat connection on $S^3$, and the partition function is given by $Z(S^3) = \sqrt{\frac{2}{\kappa +2}} \sin \left( \frac{\pi}{\kappa +2} \right)$ \cite{Witten:1988hf}. The second quotiening scheme, in which the modding out is done by a smaller space, the space of homotopically trivial gauge transformations $\mathcal G_0$. Consequently, we have a wider moduli space, containing $\mathbb Z$ many flat connections, which can be read from the homotopy relation $\Pi_3(SU(2)) = \mathbb Z$, and these flat connections are labeled by an integer which is closely related to the instanton number if one views $I \sim \int_{S^3} CS(A)$ as $c_2 \sim \int_M \tr F\wedge F$ with $\partial M = S^3$. In the below construction, we will use the latter quotiening, where we \emph{do} distinguish, physically, configurations with different winding numbers. \\

Getting back to the saddle point evaluation, if one expands $I$ around a classical point $\mathcal A(x) \in \mathscr M_{\text{CS}}$ to second order, one gets
\begin{equation} \label{Z(S3) semiclassical approximated}
    Z(S^3) \approx \frac{1}{\text{vol} (\mathcal G_0)} \int_{\mathscr M_{\text{CS}}} \mathscr D \mathcal A(x) \ e^{-I[\mathcal A(x)]} \det \Big( I''[\mathcal A(x)] \Big)^{-1/2},
\end{equation}
where we divide the path integral by $\frac{1}{\text{vol} (\mathcal G_0)}$ to implement the modding out of the gauge redundancy. We would like to show that the reduced theory captures the original one at the level of the path integral, in the semiclassical regime. To achieve this, we will argue as follows.\\

We first return to (\ref{Z(S3) semiclassical approximated}). The determinant appearing here can be written as
\begin{equation}
    \det \Big( I''[\mathcal A] \Big) = \int \mathscr D \omega \exp \left( \frac{i\kappa}{4\pi} \int \tr\ \omega \wedge d_{\mathcal A} \omega  \right),
\end{equation}
where $\omega$ is an adjoint valued 1-form and $d_{\mathcal{A}} \omega = d \omega + [\mathcal A, \omega ]$ is the gauge covariant derivative of $\omega$ with respect to the flat background $\mathcal A$. The determinant is gauge invariant, but to get a sensible answer in the evaluation, one needs gauge fixing. We will not be interested in evaluating the determinant; what we will be interested in is its invariance under $\mathcal G_0$, which we will use to recast it in a way compatible with the spherically symmetric reduction. That the determinant is invariant under the action of $\mathcal G_0$ means that it depends only on the integer $\Pi_3(S^3) = \mathbb Z$ which is the winding number of the flat connection $\mathcal A$. The flat connection can be written as $\mathcal A = g_n d g_n^{-1}$ with $g_n$ a gauge element of the winding number $n$. We can also view $g_n$ as a map from the space-time $S^3$ to the gauge group $SU(2) \cong S^3$, $g_n : S^3 \mapsto S^3$, and such mappings have integer degree given by
\begin{equation}
    \text{deg} (g) = - \frac{1}{24\pi^2} \int_{S^3} \tr \Big( gdg^{-1} \wedge gdg^{-1} \wedge gdg^{-1} \Big) \in \mathbb Z.
\end{equation}
In particular, under infinitesimal deformations $g_n \xrightarrow{} g_n + \delta g$, the degree, or the winding number, remains invariant. With these infinitesimal deformations, we can build up all the small gauge transformations in the winding sector $n$, or, equivalently, the orbit of $g_n$ under the action of $\mathcal G_0$. Because our quotiening scheme is such that we are modding out $\mathcal G_0$, the determinant does not change under such deformations. \\ 

Clearly, in addition to the determinant, the on-shell action is also invariant against infinitesimal deformations of $g_n$ or small gauge transformations. The easiest way to see this is to write $CS(A) = \tr \big( A \wedge F - \frac{1}{3} A\wedge A \wedge A \big)$ and observe that when $A$ is a flat connection $\mathcal A = gdg^{-1}$, the integral $\int_{S^3} CS(\mathcal A)$ becomes proportional to the winding number $n$ of $g$:
\begin{equation}
    -I[\mathcal A] =  \frac{i\kappa}{4\pi} \int_{S^3} CS(\mathcal A) = 2\pi i \kappa n.
\end{equation}
So, the value of $I$ depends only on the topological sector in which $\mathcal A$ lives; in fact, $e^{-I[\mathcal A]} =1$ due to the quantization of $\kappa$. Therefore, the integral in (\ref{Z(S3) semiclassical approximated}) reduces to a discrete sum over integers, where for each $n$, the on-shell action and the determinant are evaluated on an arbitrary 
\footnote{It is arbitrary because if we start with $g_n$, and deform that with a small gauge transformation $g_n \xrightarrow{} g_n + \delta g$, the summand will not change.} flat connection of degree $n$. We denote the space of flat connections of winding number $n$ as $\mathscr M^n_{\text{CS}}$, choose a representative flat connection in that winding sector, call it $\mathcal A_n \in \mathscr M_{\text{CS}}^n$, and write (\ref{Z(S3) semiclassical approximated}) as:
\begin{equation} \label{Semiclassical evaluation as a discrete sum over n}
\begin{aligned}
    Z(S^3) &\approx \frac{1}{\text{vol} (\mathcal G_0)} \sum_{n=-\infty}^{\infty} \int_{\mathscr M_{\text{CS}}^n } \mathscr D \mathcal A_n e^{-I[\mathcal A_n]} \det \Big( I''[\mathcal A_n] \Big)^{-1/2} \\
    &= \frac{1}{\text{vol} (\mathcal G_0)} \sum_{n = -\infty}^{\infty} e^{-I[\mathcal A_n]} \det \Big( I''[\mathcal A_n] \Big)^{-1/2} \text{vol}(\mathscr M_{\text{CS}}^n ),
\end{aligned}
\end{equation}
where $\text{vol}(\mathscr M_{\text{CS}}^n ) = \int_{\mathscr M_{\text{CS}}^n} \mathscr D \mathcal A_n$ is the volume of the space of flat connections of degree $n$. We note that the moduli space of flat connections of degree $n$ is given by $\mathscr M_{\text{CS}}^n / \mathcal G_0$, which has only one element in it. Consequently, $\frac{\text{vol}(\mathscr M_{\text{CS}}^n )}{\text{vol} (\mathcal G_0)}$ = 1 and therefore, that factor disappears from the sum. For the rest of the summand, we can choose any flat connection $\mathcal A_n$ we want as long as it has the correct degree. Now, here comes the key observation in showing that the 1d reduced theory captures the path integral of the $S^3$ Chern-Simons theory. We can always deform $\mathcal A_n$ in the summand to make it spherically symmetric, in the sense that it can be written in the form (\ref{Spherically symmetric ansatz on A}). As such, we can write $Z(S^3)$ as
\begin{equation} 
    Z(S^3) \approx \sum_{n=-\infty}^{\infty} e^{-I[\mathcal A_n(r)]} \det \Big( I''[\mathcal A_n(r)] \Big)^{-1/2},
\end{equation}
where $\mathcal A_n(r)$ is a spherically symmetric flat connection of degree $n$. We denote the space of spherically symmetric flat connections of degree $n$ with $\mathscr M_r^n$; insert 1 into the above sum in the following way
\begin{equation} \label{Z(S^3) = sum e det 1/vol(Mr) int_(Mr)}
    Z(S^3) \approx \sum_{n=-\infty}^{\infty} e^{-I[\mathcal A_n(r)]} \det \Big( I''[\mathcal A_n(r)] \Big)^{-1/2}  \frac{1}{\text{vol}(\mathscr M_r^n )} \int_{\mathscr M_r^n} \mathscr D \mathcal A_n(r).
\end{equation}
Now, one can take the exponential and the determinant inside the integral over $\mathscr M_r^n$, because they are invariant under infinitesimal deformations $\delta g_r \in \mathcal G_0^r$ that leaves $g_n$ radially symmetric, which form a subspace of all small gauge transformations $\mathcal G_0^r \subset \mathcal G_0$:
\begin{equation}
    Z(S^3) \approx \sum_{n=-\infty}^{\infty} \frac{1}{\text{vol}(\mathscr M_r^n )} \int_{\mathscr M_r^n} \mathscr D \mathcal A_n(r)  e^{-I[\mathcal A_n(r)]} \det \Big( I''[\mathcal A_n(r)] \Big)^{-1/2}.
\end{equation}
If the volume of the solution space of spherically symmetric flat connections $\mathscr M_r^n$ is independent of $n$ $-$such is the case for $\mathscr M_{\text{CS}}^n$ because each sector is generated by the orbit of $g_n \in \mathscr M_{\text{CS}}^n$ under the action of $\mathcal G_0-$ then one can take that factor out of the sum over $n$. To see that this is the case, one can observe that by taking a radially symmetric gauge element of degree $n$: $g_n(r)$ and acting on it with the group of all radially symmetric gauge transformations $\mathcal G_0^r$, we generate $\mathscr M_r^n$. Each winding sector is generated by $\mathcal G_0^r$, which is independent of $n$. It follows that $\text{vol}(\mathscr M_r^n)$ is independent of $n$, and is equal to $\text{vol}(\mathcal G_0^r)$. One can also show that $\mathcal G_0^r$ is isomorphic to $U(1)$, which can be seen from section \ref{sec 2: Euclidean SU(2) CS Theory on S1 * S2} as the reduced theory under the spherically symmetric ansatz gives a $U(1)$ gauge theory. The one-parameter subgroup of $SU(2)$ given by $U = \exp \big( f(r) \frac{x^a}{r} t^a \big)$ comprises precisely the elements of $\mathcal G_0^r$. From this, it follows that
\begin{equation} \label{Z approx int Mr e det}
\begin{aligned}
    Z(S^3) &\approx \frac{1}{\text{vol} (\mathcal G_0^r)} \sum_{n=-\infty}^{\infty} \int_{\mathscr M_r^n} \mathscr D \mathcal A_n(r)  e^{-I[\mathcal A_n(r)]} \det \Big( I''[\mathcal A_n(r)] \Big)^{-1/2} \\
    &= \frac{1}{\text{vol} (\mathcal G_0^r)} \int_{\mathscr M_r} \mathscr D \mathcal A(r) e^{-I[\mathcal A(r)]} \det \Big( I''[\mathcal A(r)] \Big)^{-1/2},
\end{aligned}
\end{equation}
where we combined the integral over the $n^{\text{th}}$ sector and the discrete sum into a single integral over the full solution space of spherically symmetric flat connections $\mathscr M_r$. Now we argue that the right-hand side is the semiclassical evaluation of the path integral of Chern-Simons theory restricted to the spherically symmetric field configurations. That is to say, the semiclassical evaluation of 
\begin{equation} 
    Z_{\text{spherical}} \equiv \frac{1}{\text{vol} (\mathcal G_0^r) } \int_{\text{spherical}} \mathscr D A(r) e^{-I[A(r)]} ,
\end{equation}
with the field space restricted to configurations that can be written in the form (\ref{Spherically symmetric ansatz on A}), gives 
\begin{equation} \label{Zspherical = int Mr e det}
    Z_{\text{spherical}} \approx \frac{1}{\text{vol} (\mathcal G_0^r)} \int_{\mathscr M_r} \mathscr D \mathcal A(r) e^{-I[\mathcal A(r)]} \det \Big( I''[\mathcal A(r)] \Big)^{-1/2}.
\end{equation}
Here, $A(r)$ is a gauge field configuration that can be written in the form (\ref{Spherically symmetric ansatz on A}) but is not necessarily a solution to the equations of motion. On the other hand $\mathcal A(r)$ is a spherically symmetric flat gauge field, so it is in $\mathscr M_r$. Clearly, the full solution space of the restricted path integral is given by $\mathscr M_r$, because the main integral is restricted only to the spherically symmetric configurations. We can decompose the integral over $\mathscr M_r$ into a discrete sum over the winding sectors $n$, and in each sector we have the integral of $e^{-I[\mathcal A_n(r)]} \det(I''(\mathcal A_n(r))^{-1/2}$ over $\mathscr M_r^n$. Thus, a comparison of (\ref{Z approx int Mr e det}) and (\ref{Zspherical = int Mr e det}) shows that
\begin{equation} \label{Z(S^3) = Zspherical}
    Z(S^3) \cong Z_{\text{spherical}},
\end{equation}
where, by $\cong$ we mean the semiclassical limit of the Chern-Simons theory can be evaluated by constraining the configuration space to the fields of the form (\ref{Spherically symmetric ansatz on A}). We may expect a relation between $Z_{\text{spherical}}$ and the partition function of the reduced theory (\ref{Path integral of the S3 reduced theory}), because for spherically symmetric configurations of the form (\ref{Spherically symmetric ansatz on A}), the Chern-Simons action reduces to an action on $I_\infty$. Moreover, the measure $\mathscr D A(r)$ may also be related to the measure $\mathcal D A \mathcal D Y \mathcal D \overline Y$. So, one would expect a relationship between the two
\begin{equation}
    Z_{\text{spherical}} \simeq Z(S^3 \xrightarrow{} I_\infty),
\end{equation}
in the sense that the right-hand can be seen as an effective description of the left-hand side. One may object to this based on the fact that we obtained $Z(S^3 \to I_\infty) = \kappa^{-1}$ and the large $\kappa$ behavior of $S^3$ Chern-Simons theory with $SU(2)$ gauge group was found $Z(S^3) \sim \kappa^{-3/2}$ in \cite{Witten:1988hf}. However, one must keep in mind that in our quotiening, there are $\mathbb Z$ many elements in the moduli space of flat connections; whereas in \cite{Witten:1988hf} the quotiening is such that there is only one element in the moduli space. Different quotients of the gauge field space lead to different configuration spaces, and hence to different moduli spaces, and this changes the physics of the theory. Our construction is based on the configuration space $\mathcal C = \mathcal A / \mathcal G_0$, whereas that of \cite{Witten:1988hf} is based on $\mathcal C' = \mathcal A / \mathcal G$, and here $\mathcal A$ denotes the space of all $\mathfrak{su}(2)$-valued connections on $S^3$. There is no reason to expect that two theories, defined on different configuration spaces $\mathcal C$ and $\mathcal C'$, to exactly agree on their partition function. \\

Having established a semiclassical connection between $Z(S^3)$ and $Z_{\text{spherical}}$, we now discuss the quantum observables in both theories and their possible relation. We can generalize the above argument that led to (\ref{Z(S^3) = Zspherical}) for the functional integral of any gauge invariant functional $\mathcal F[\mathcal A]$ over the space of flat connections. One has
\begin{equation}
\begin{aligned}
    \frac{1}{\text{vol}(\mathcal G_0)} \int_{\mathscr M_{\text{CS}}} \mathscr D \mathcal A \mathcal F[\mathcal A] &= \frac{1}{\text{vol}(\mathcal G_0)}  \sum_{n=-\infty}^{\infty} \int_{\mathscr M_{\text{CS}}^n} \mathscr D \mathcal{A}_n \mathcal F[\mathcal A_n] = \sum_{n = -\infty}^{\infty} \mathcal{F}[\mathcal{A}_n] \\
    \frac{1}{\text{vol}(\mathcal G_0^r)} \int_{\mathscr M_{r}} \mathscr D \mathcal A(r) \mathcal F[\mathcal A(r)] &= \frac{1}{\text{vol}(\mathcal G_0^r)} \sum_{n=-\infty}^{\infty} \int_{\mathscr M_{r} ^n} \mathscr D \mathcal{A}_n(r) \mathcal F[\mathcal A_n(r)] = \sum_{n = -\infty}^{\infty} \mathcal{F}[\mathcal{A}_n(r)] .
\end{aligned}
\end{equation}
Because $\mathcal F[A_n]$ is invariant under infinitesimal deformations, one can always find a small gauge transformation to make it spherically symmetric without changing the summand. This is essentially the same argument that we made above for the special case of $\mathcal F[\mathcal A] = e^{-I[\mathcal A]} \det \big( I''[\mathcal A] \big)^{-1/2}$. 
Consequently, we can show that the two lines give the same results for a general gauge invariant functional $\mathcal F[\mathcal A]$. 
\begin{equation}
    \frac{1}{\text{vol}(\mathcal G_0) } \int_{\mathscr M_{\text{CS}}} \mathscr D \mathcal A \mathcal F[\mathcal A] = \sum_{n=-\infty}^{\infty} \mathcal{F}[\mathcal A_n] = \frac{1}{\text{vol}(\mathcal G_0^r)} \int_{\mathscr M_r } \mathscr D \mathcal A(r) \mathcal F[\mathcal A(r)].
\end{equation}
The two theories have different gauge groups and live in different spacetime dimensions, yet they agree on the semiclassical level \footnote{ What we did can be seen as hand-picking the space of all flat connections compatible with spherical symmetry, and upon changing how we gauge the theory in accordance with the symmetry reduction, we can reduce the theory while having the same quantum observables semiclassically.}. That this equality holds for arbitrary gauge invariant functionals implies that the quantum observables of the 3-dimensional theory agree, in the large $\kappa$ limit, with the quantum observables of the 1-dimensional theory, because the observables are gauge invariant. In Chern-Simons gauge theory, the only gauge invariant observables are the Wilson loop operators, defined by
\begin{equation}
    W_R(C) \equiv \tr_R P \exp \Big( \oint_C A \Big),
\end{equation}
with $\tr_R$ the trace in the representation $R$, $P$ is the path ordering operator, and $C$ is a curve over which the extended operator is defined. The correlation function of some product of Wilson lines is defined as
\begin{equation}
    \Big \langle \prod_{i} W_{R_i}(C_i) \Big \rangle = \int \mathscr D A \ e^{-I[A]} \prod_{i} W_{R_i}(C_i).
\end{equation}
Since $W_R(C)$ is gauge invariant, in the large $\kappa$ limit we expect these correlation functions to reduce to the correlation functions of the 1-dimensional theory.
Finding the corresponding observable to the Wilson operators, in the reduced theory, is an open problem.  \\

The conclusion is that the semiclassical evaluation of the quantum Chern-Simons theory can be dealt with by considering only the classical solutions that possess spherical symmetry. The partition function and the observables of both theories agree in the semiclassical regime. We summarize the results in Table \ref{Table 1}.

\begin{center} \label{Table 1}
\begin{table}  
\begin{tabular}{ccc}
\multicolumn{1}{l}{} &
  \multicolumn{1}{l}{} &
  \multicolumn{1}{l}{} \\ \hline
\multicolumn{1}{|c|}{\multirow{2}{*}{$\kappa \xrightarrow{} \infty$}} &
  \multicolumn{1}{c|}{\multirow{2}{*}{\begin{tabular}[c]{@{}c@{}}Chern-Simons Theory   \\ on $S^3$ \end{tabular}}} &
  \multicolumn{1}{c|}{\multirow{2}{*}{\begin{tabular}[c]{@{}c@{}}Symmetry Reduced Chern-Simons \\ Theory on $I_\infty=[0,\infty]$ \end{tabular}}} \\
\multicolumn{1}{|c|}{} &
  \multicolumn{1}{c|}{} &
  \multicolumn{1}{c|}{} \\ \hline
\multicolumn{1}{|c|}{\multirow{2}{*}{Action}} &
  \multicolumn{1}{c|}{\multirow{2}{*}{$I[A]$}} &
  \multicolumn{1}{c|}{\multirow{2}{*}{$S_{\text{CS}}[A,Y,\overline Y]$}} \\
\multicolumn{1}{|c|}{} &
  \multicolumn{1}{c|}{} &
  \multicolumn{1}{c|}{} \\ \hline
\multicolumn{1}{|c|}{\multirow{2}{*}{Partition Function}} &
  \multicolumn{1}{c|}{\multirow{2}{*}{$Z = \sum_n \det \Big( I''[\mathcal A] \Big)^{-1/2}$}} &
  \multicolumn{1}{c|}{\multirow{2}{*}{$Z_{\text{spherical}}  =\sum_n \det \Big( I''[\mathcal A(r)] \Big)^{-1/2} \cong Z$}} \\
\multicolumn{1}{|c|}{} &
  \multicolumn{1}{c|}{} &
  \multicolumn{1}{c|}{} \\ \hline
\multicolumn{1}{|c|}{\multirow{2}{*}{Observables}} &
  \multicolumn{1}{c|}{\multirow{2}{*}{$W_R(C)$}} &
  \multicolumn{1}{c|}{\multirow{2}{*}{$?$}} \\
\multicolumn{1}{|c|}{} &
  \multicolumn{1}{c|}{} &
  \multicolumn{1}{c|}{} \\ \hline
\end{tabular}
\caption{The full and the reduced theory}
\end{table}
\end{center}

\subsection{Generalization to \texorpdfstring{$\mathcal{S}ym$}{Sym}-Reduction of Quantum Chern-Simons Theory on \texorpdfstring{$S^3$}{S**3} with Gauge Group \texorpdfstring{$G$}{G} } \label{subsec 3.2: SYM reduction of G CS on S^3}

We generalize the result $Z(S^3) \cong Z_{\text{spherical}}$ in two directions. Our logical development here is in parallel with that of the previous subsection for a smoother read.\\

The first line of generalization is changing the gauge group from $SU(2)$ to a general simple Lie group $G$. Due to Bott's remarkable theorem \cite{Coleman:1978ae}, for a general simple Lie group, we have $\Pi_3(G) = \mathbb Z$, which is a result familiar from the Yang-Mills instanton literature \cite{Shifman:1994ee}. With such a group, (\ref{Semiclassical evaluation as a discrete sum over n}) is still the same, but now $\mathcal A_n$ is a flat connection valued in the lie algebra $\mathfrak g$, and we denote the space of flat connections of degree $n$ associated to the group $G$ as $\mathscr M_G^n$. So, (\ref{Semiclassical evaluation as a discrete sum over n}) reads 
\begin{equation} 
    Z(S^3,G) \approx \frac{1}{\text{vol} (\mathcal G_0) } \sum_{n=-\infty}^{\infty} e^{-I[\mathcal A_n]} \det \Big( I''[\mathcal A_n] \Big)^{-1/2} \text{vol}(\mathscr M_G^n) ,
\end{equation}
with $\mathcal G_0$ being the space of homotopically trivial gauge transformations that are elements of $G$, and $\mathcal A_n \in \mathscr M_G^n$. Just as in the $SU(2)$ case, the moduli space of flat connections of degree $n$ is given by $\mathscr M_G^n / \mathcal G_0$, which has one element. Thus, $Z(S^3,G)$ in the semiclassical limit reads as
\begin{equation} \label{Z(S^3,G) = sum e 1/sqrt(det) }
    Z(S^3,G) \approx \sum_{n=-\infty}^{\infty} e^{-I[\mathcal A_n]} \det \Big( I''[\mathcal A_n] \Big)^{-1/2}.
\end{equation}
Another line of generalization is the symmetry reduction scheme. For gauge fields valued in the lie algebra $\mathfrak g$, one may cook up new symmetry reduction schemes, and as long as that symmetry is associated with a compact group, we can invoke Palais' symmetric criticality to obtain a reduced action \footnote{If the symmetry is associated with a non-compact group, the integration over the symmetry parameters would give infinity so the reduced action would not be sensible. For example, one common reduction scheme associated to a non-compact group is the translational symmetry along a particular axis, say $x_s$. If all the fields in the theory are imposed to be independent of $x_s$, then one could integrate over $x_s$ in the action, and if it is a non-compact coordinate the integration yields $\infty$, so the reduced action is ill-defined. }. We abstractly denote such a symmetry reduction scheme as $\mathcal{S}ym$, and with that scheme, one can obtain a reduced theory whose partition function agrees with that of $Z(S^3, G)$ in the large $\kappa$ limit. Let us verify this. We denote a gauge element of degree $n$, $g_n(x) \in \mathcal G_n$ that is compatible with the symmetry ans\"atze, as $(g_{\mathcal{S}ym})^n$. With such a gauge element, we can construct a flat connection of winding number $n$: $(\mathcal A_{\mathcal{S}ym})_n = (g_{\mathcal{S}ym})^n d (g_{\mathcal{S}ym}^{n})^{-1}$. Thus, (\ref{Z(S^3,G) = sum e 1/sqrt(det) }) can be written as
\begin{equation}
    Z(S^3,G) \approx \sum_{n=-\infty}^{\infty} e^{-I[(\mathcal A_{\mathcal{S}ym} )_n]} \det \Big( I''[(\mathcal A_{\mathcal{S}ym} )_n] \Big)^{-1/2}.
\end{equation}
In this sum, we insert 1 analogously as in (\ref{Z(S^3) = sum e det 1/vol(Mr) int_(Mr)})
\begin{equation}
    Z(S^3,G) \approx \sum_{n=-\infty}^{\infty} e^{-I[(\mathcal A_{\mathcal{S}ym} )_n]} \det \Big( I''[(\mathcal A_{\mathcal{S}ym} )_n] \Big)^{-1/2} \frac{1}{\text{vol}(\mathscr M_{{\mathcal{S}ym}}^n) } \int_{\mathscr M_{\mathcal{S}ym} ^n } \mathscr D (\mathcal A_{\mathcal{S}ym} )_n,
\end{equation}
where we denoted the space of flat connections \& gauge transformations of winding number $n$, compatible with the symmetry reduction ${\mathcal{S}ym}$, as $\mathscr M_{\mathcal{S}ym}^n$ \& $\mathcal G_{\mathcal{S}ym}^n$, respectively. Clearly, $\text{vol}(\mathscr M_{\mathcal{S}ym}^n) = \text{vol} (\mathcal G_{\mathcal{S}ym}^n)$. Moreover, $\mathcal G_{\mathcal{S}ym}^n$ can be generated by the orbit of a particular $g_{\mathcal{S}ym} \in \mathcal G_{\mathcal{S}ym}^n$ under the action of small gauge transformations compatible with ${\mathcal{S}ym}$: $\mathcal G_{\mathcal{S}ym}^0$. Therefore, the volumes are independent of $n$: $\text{vol}(\mathscr M_{\mathcal{S}ym}^n) = \text{vol} (\mathcal G_{\mathcal{S}ym}^n) = \text{vol} (\mathcal G_{\mathcal{S}ym}^0)$. We can also take the exponential and the determinant inside the integral because they are invariant against infinitesimal deformations compatible with ${\mathcal{S}ym}$, which is a subset of the space of all infinitesimal deformations $\mathcal G_{\mathcal{S}ym}^0 \subset \mathcal G_0$. Therefore, $Z$ reads
\begin{equation} \label{Z(G,SYM) = int M(SYM) e 1/sqrt(det) }
\begin{aligned}
    Z(S^3,G) &\approx \frac{1}{\text{vol}(\mathcal G_{\mathcal{S}ym}^0) } \sum_{n=-\infty}^{\infty} \int_{\mathscr M_{\mathcal{S}ym}^n} \mathscr D (\mathcal A_{\mathcal{S}ym})_n e^{-I[(\mathcal A_{\mathcal{S}ym} )_n]} \det \Big( I''[(\mathcal A_{\mathcal{S}ym} )_n] \Big)^{-1/2}  \\
    &= \frac{1}{\text{vol}(\mathcal G_{\mathcal{S}ym}^0) }  \int_{\mathscr M_{\mathcal{S}ym}} \mathscr D \mathcal A_{\mathcal{S}ym} e^{-I[\mathcal A_{\mathcal{S}ym} ]} \det \Big( I''[\mathcal A_{\mathcal{S}ym} ] \Big)^{-1/2}.
\end{aligned}
\end{equation}
We combined the discrete sum and the integral over the winding sectors to a full integral. Just as for the $G=SU(2)$ and ${\mathcal{S}ym}$ = spherical symmetry case, we can show that the path integral constrained to field configurations compatible with ${\mathcal{S}ym}$
\begin{equation}
    Z(G,{\mathcal{S}ym}) \equiv \int_{\mathcal{S}ym} \mathscr D A_{\mathcal{S}ym} e^{-S_{\mathcal{S}ym} [A_{\mathcal{S}ym}] },
\end{equation}
is semiclassically the same as (\ref{Z(G,SYM) = int M(SYM) e 1/sqrt(det) }):
\begin{equation}
    Z(G, {\mathcal{S}ym} ) \underbrace{\approx}_{\text{large }\kappa \text{ limit}} \int_{\mathscr M_{\mathcal{S}ym} } \mathscr D \mathcal A_{\mathcal{S}ym} e^{-I[\mathcal A_{\mathcal{S}ym}] } \det \Big( I''[\mathcal A_{\mathcal{S}ym} ] \Big)^{-1/2}.
\end{equation}
$A_{\mathcal{S}ym}$ is a field configuration compatible with ${\mathcal{S}ym}$, but is not necessarily a flat connection; whereas $\mathcal A_{\mathcal{S}ym}$ is a ${\mathcal{S}ym}$-compatible flat connection, so it is in $\mathscr M_{\mathcal{S}ym}$. Finally, we have the agreement
\begin{equation}
    Z(S^3,G) \cong Z(G,{\mathcal{S}ym} )
\end{equation}
in the large $\kappa$ limit. Replacing $e^{-I} \det(I'')^{-1/2}$ with any gauge invariant functional $\mathcal F[\mathcal A]$ will give the same result:
\begin{equation}
    \frac{1}{\text{vol}(\mathcal G_0)} \int_{\mathscr M_{G}} \mathscr D \mathcal A \mathcal F[\mathcal A] = \sum_{n=-\infty}^{\infty} \mathcal F[\mathcal A_n] = \frac{1}{\text{vol} (\mathcal G_{\mathcal{S}ym}^0)} \int_{\mathscr M_{\mathcal{S}ym}} \mathscr D (\mathcal A_{\mathcal{S}ym} )_n \mathcal F [ (\mathcal A_{\mathcal{S}ym} )_n] .
\end{equation}
It is reasonable to expect that $Z(G, {\mathcal{S}ym})$ can effectively be described by a reduced theory, depending on the reduction scheme ${\mathcal{S}ym}$. One may try different ${\mathcal{S}ym}$-reduction schemes for arbitrary groups $G$, and it will agree with $Z(S^3, G)$ in the large $\kappa$ limit. With specific choices of $\mathcal F$, one might relate the quantum observables of the original theory with the ${\mathcal{S}ym}$-reduced theory. \\

Up to here, we had two lines of generalizations. The first is generalizing the group $G$ and the second is generalizing the symmetry reduction scheme from spherical symmetry to an abstract scheme that we denoted ${\mathcal{S}ym}$. One may be interested in constructing similar dualities for Chern-Simons theories on an arbitrary topology other than $S^3$, which may allow more interesting symmetry reduction schemes\footnote{For example, as we discuss in section \ref{sec 5: Disk Reduction of Chern-Simons Theories on X = R * D2}, for topologies $X = \mathbb R \times D_2$, there is a symmetry reduction on the disk that yields 2-dimensional $BF$-theories.}. However, $\Pi_3(G) = \mathbb Z$ is an important step in the argument, allowing us to break integrals into discrete sums, and such a thing may not be viable for a general manifold. \\

We would like to reflect on the importance of the group structure in the solution space of the Chern-Simons theory. This allowed us to evaluate the summand (\ref{Z(S^3,G) = sum e 1/sqrt(det) }) in terms of field configurations having specific symmetry conditions, without changing the result. This is because the solution space $\mathscr M_G^n$ has a group structure and one can always take an arbitrary element in this space $\mathcal A_n \in \mathscr M_G^n$ and deform it to a configuration $(\mathcal A_{\mathcal{S}ym} )_n$ still in $\mathscr M_G^n$, while $Z$ remains invariant because this deformation corresponds to a gauge transformation that we mod out. Therefore, we can hand-pick certain regions of the space of flat connections based on a symmetry condition and, as long as we change the gauge group compatible with that symmetry, we get reduced Quantum Field Theories dual to the original theory in the semiclassical regime. For example, in the $G=SU(2)$, ${\mathcal{S}ym}$ = spherical symmetry case, we hand-picked the flat connections possessing spherical symmetry, and we have morphed the $SU(2)$ gauge symmetry to a $U(1)$ gauge symmetry, in doing so obtaining the same theory semiclassically\footnote{This holds semiclassically because in that regime the critical points of the action dominate, and due to the group structure, we can arrange the saddle-points such that all of them are compatible with the $\mathcal{S}ym$-reduction. Quantum corrections at loop orders may or may not spoil this correspondence.}. This is a neat result, coming from the group structure in the solution space. One can generate families of theories, reduced from Chern-Simons theories, which are dual to each other in the sense that their quantum observables (including the identity observer that is the partition function) agree in the semiclassical limit. With another theory having space of solutions that has a group structure, one can generate the same results by repeating the lines with some relatively straightforward modifications.

\section{\texorpdfstring{$SL(2,\mathbb{C})$}{SL(2,C)} Chern-Simons Theory or de Sitter Gravity} \label{sec 4: SL(2,C) Chern Simons or dS_3 gravity}
For $SL(2,\mathbb{C})$ Chern-Simons gauge theory, the connection is a doublet of the form $(A, \bar A ) = ( A^a t^a, \bar A^a t^a)$. The action is given by (we are using conventions from \cite{Verlinde:2024zrh, Witten:1989ip}) :
\begin{equation}
    I = \frac{t}{4\pi} \int_{\mathcal M} CS(A) - \frac{\overline t}{4\pi} \int_{\mathcal M} CS(\bar A).
\end{equation}
This action is equivalent to the action of de Sitter Gravity in the first-order formalism if one makes the identifications:
\begin{equation}
    \kappa = \frac{1}{4G_N} \quad ; \quad A^a = \omega^a + i e^a \quad ; \quad \bar A^a = \omega^a - i e^a ,
\end{equation}
where $e^a$ is the dreibein and $\omega^a$ is the spin connection. The field equations are solved by flat connections $F=0$ and $\bar F = 0$. On the dS$_3$ gravity side, the field equations give Einstein's equations with a positive cosmological constant (with the radius of de Sitter space taken to be 1) in the first-order formalism. In the gauge theory side, let us employ the spherically symmetric ansatz 
\begin{equation}
    \begin{aligned}
    A &= t^a \Bigg[ \varepsilon_{iak} \frac{x_k}{r^2} (\varphi_2 - 1) + \frac{\delta^{\perp}_{ia}}{r} \varphi_1  +  \frac{x_ix_a}{r^2} A \Bigg] dx^i, \\
    \bar A &= t^a \Bigg[ \varepsilon_{iak} \frac{x_k}{r^2} (\overline\varphi_2 - 1)  + \frac{\delta^{\perp}_{ia}}{r} \overline\varphi_1 - \frac{x_ix_a}{r^2} \overline A \Bigg] dx^i.
    \end{aligned}
\end{equation}
We will set $t = i\kappa$ and $\overline t = -i\kappa$. After reducing the action, and dropping the boundary terms $\int dr \partial_r\varphi_1$ and $\int dr \partial_r\overline \varphi_1$, we get: 
\begin{equation}
    S_{\text{CS}} = \kappa I[A,\varphi_1,\varphi_2]  - \kappa I[\overline A, \overline \varphi_1, \overline\varphi_2]
\end{equation}
where
\begin{equation} 
    \begin{aligned}
        I[A,\varphi_1,\varphi_2] = i \int dr \Big( \varphi_1 \varphi_2' - \varphi_1' \varphi_2 + A (\varphi_2^2 + \varphi_1^2 - 1) \Big). 
    \end{aligned}
\end{equation}
We define the complex scalar fields 
\begin{equation}
    \begin{aligned}
        Y_1 = \varphi_1 + i \varphi_2 \quad &; \quad Y_2 = \overline \varphi_1 - i \overline \varphi_2, \\
        \overline Y_1 = \varphi_1 - i \varphi_2 \quad &; \quad \overline Y_2 = \overline \varphi_1 + i \overline \varphi_2,
    \end{aligned}
\end{equation}
in terms of which the reduced action reads
\begin{equation} \label{Reduced action of SL(2,C) CS in terms of A,Y, bar A, bar Y}
\begin{aligned}
    S_{\text{CS}} = \kappa \int dr \Big( \overline Y_1 (\partial_r + iA) Y_1 - i A \Big) 
    + \kappa \int dr \Big( \overline Y_2 (\partial_r + iA) Y_2 - i \overline A \Big) .
\end{aligned}
\end{equation}
As we did in section \ref{sec 2: Euclidean SU(2) CS Theory on S1 * S2}, we can map the classical solutions of this theory to a fermionic theory, which we do not repeat here. \\ 

Based on these results, we can conclude that through symmetry reduction of the 3-dimensional dS-gravity, one gets two copies of a 1-dimensional theory that has a fermionic character. This fermionic character is present not only at the classical level, one can recast the partition function based on the action (\ref{Reduced action of SL(2,C) CS in terms of A,Y, bar A, bar Y}) as the partition function of a theory with two complex bosons and two complex fermions, a boson-fermion pair for each field $Y_1, Y_2$. 

\subsection{Semiclassical Limit of \texorpdfstring{$SL(2,\mathbb C)$}{SL(2,C)} Quantum Chern-Simons Theory} \label{subsec 4.1: Semiclassical Limit of SL(2,C) Chern-Simons Gauge Gheory}

Let us study the large $\kappa$ limit \cite{Witten:1989ip} of $SL(2,\mathbb C)$ Chern-Simons theory on $\mathcal M = S^3$. We will use some conclusions from section \ref{sec 3: Symmetry Reduction at the Semiclassical Regime}. The path integral is given as
\begin{equation}
    Z = \int \mathscr D  A \mathscr D \bar A \exp\Bigg\{ \frac{i\kappa}{4\pi} \int_{\mathcal M} CS( A) - \frac{i\kappa}{4\pi} \int_{\mathcal M} CS(\bar A) \Bigg\}.
\end{equation}
Based on the results of section \ref{sec 2: Euclidean SU(2) CS Theory on S1 * S2} and section \ref{sec 3: Symmetry Reduction at the Semiclassical Regime}, we have
\begin{equation}
    Z \cong Z_{\text{spherical}} \times \overline Z_{\text{spherical}},
\end{equation}
with $Z_{\text{spherical}}$ and $\overline Z_{\text{spherical}}$ partition functions of which the measures are constrained to the spherically symmetric field configurations. The fields obey boundary conditions depending on which topology we choose for $\mathcal M$, periodic for $S^1\times S^2$ and twisted for $S^3$. 

\subsection{Possible Applications} \label{subsec 4.2: possible applications}
In three-dimensional gravity, a spherically symmetric ansatz in the first-order formalism was employed and studied by one of the authors \cite{Tekin:1998wu, Tekin:1999gn, Ferstl:2000qp}. In \cite{Tekin:1999gn}, the relevance of the singular solutions, which are needed in the canonical quantization, was demonstrated in the path-integral approach using an ansatz on the dreibein and the spin-connection which is analogous to our ansatz on the gauge fields in the present paper. It would be illuminating to study the saddle points of the $3d$ gravity with such a reduction and possibly relate it to the results on the gauge theory side. Another place where this reduction may be fruitful to study is in understanding the relation between dS$_3$ gravity and SYK models better, which is under study \cite{Verlinde:2024zrh, Verlinde:2024znh, Narovlansky:2023lfz, Susskind:2022bia}. The gauge theory formalism of de Sitter gravity seems to have some connections with the double-scaled SYK model (DSSYK). It would be interesting to establish a connection between DSSYK and the reduced theory we obtained from $SL(2,\mathbb C)$ Chern-Simons theory. 

\section{Symmetry Reductions of Chern-Simons Theories on \texorpdfstring{$X = \mathbb R \times D_2$}{X = R * D2} to 2d BF-Theories} \label{sec 5: Disk Reduction of Chern-Simons Theories on X = R * D2} 

The general theme of the paper so far has been the reduction from 3-dimensions to 1-dimension through a monopole-soliton type spherically symmetric ansatz on the gauge potential for a Euclidean Chern-Simons theory. This section is a quick exposition of some interesting results involving Chern-Simons theories on the disk topology. Upon a symmetry reduction on the disk, we obtain $BF$-type reduced actions, which are known to be related to boundary conformal field theories (CFTs) \cite{Kapustin:2010if, Witten:1988hf, Elitzur-Shmuel-Moore-Schwimmer-Seiberg:1989nr, Tong:2016kpv, Dijkgraaf:1989pz}. The discussions here are only introductory, we shall develop the ideas presented in this section in more detail in future work. \\

In doing a symmetry reduction on the disk topology $-$we will refer to such reductions from 3d to 2d as disk reductions$-$, we use the conformal equivalence $D_2 - \{ 0 \} \sim \mathbb R^+_R \times S^1$ with $\mathbb R^+_R = (0, R) \equiv I_R$. We can see this from the metric on the disk by making a comparison with (\ref{conformal equivalence of R^3 metric with R+ * S^2 metric}), by replacing the angular element $d\Omega$ with $d\theta^2$, where $\theta$ is the angle in the polar coordinates representation of the disk. The ans\"atze that we will employ in this section will be based on this conformal equivalence. We will impose the gauge fields to be independent of the $S^1$ part (the angular coordinate), upon which the component $A_{\theta}$ acts like a scalar field that we will denote as $\phi$. The other component on the disk, $A_r$, combined with $A_t$ where $t\in \mathbb R$ makes up a $U(1)$ connection on a two-dimensional spacetime. In the case of a non-abelian disk reduction, there would be additional fields. For the $SU(2)$ disk reduction, which we discuss in subsection \ref{subsec 6.2: Disk reduction of SU(2) CS}, we get a $U(1)\times U(1)$ $BF$-theory on a two-dimensional spacetime. Moreover, there is good reason to believe that the results presented in this section have connections to the Abelianisation of Chern-Simons Theories on circle bundles of $\Sigma$ \cite{Blau:1993tv, Blau:2006gh, Beasley:2005vf}. 

\subsection{\texorpdfstring{$U(1)$}{U(1)} Chern-Simons Theory on a Disk}

It is possible to consider the abelian Chern-Simons theory under a symmetry reduction. We start with
\begin{equation} \label{U(1) CS action on X}
    I = \frac{\kappa}{4\pi} \int_X A \wedge dA,
\end{equation}
on a manifold $X = \mathbb R \times D_2$ with $D_2$ a disk with radius $R$. We will consider field configurations independent of the angular variable on the disk, the most general field configuration is
\begin{equation}
\begin{aligned}
    A_t &= A_0(r,t), \\
    A_i &= \varepsilon_{ij} \frac{x_j}{r^2} \phi(r,t) + \frac{x_i}{r} A_1(r,t), 
\end{aligned}
\end{equation}
where $r^2 = (x_ix_i)^{1/2}$ and $x_i \in D_2 \ (i=1,2)$ whereas $t \in \mathbb R$. Under a $U(1)$ gauge transformation by $\alpha(r,t)$, the change in the ansatz fields is given by
\begin{equation}
\begin{aligned}
    \phi &\xrightarrow{} \phi , \\
    A_0 &\xrightarrow{} A_0 + \partial_0 \alpha , \\
    A_1 &\xrightarrow{} A_1 + \partial_r \alpha.
\end{aligned}
\end{equation}
So $\phi$ is a gauge invariant scalar under the group of reduced $U(1)$ transformations that is of the form $e^{i\alpha(r,t)}$, whereas the $(A_0, A_1)$ pair forms a $U(1)$ potential in the $(r,t)$ half-plane. Inserting the ansatz into the action and integrating over the angle, one gets
\begin{equation}
    S_{\text{CS}} = \frac{\kappa}{2} \int dr dt \ \Big( \phi (\partial_r A_0 - \partial_0 A_1) + A_1 \partial_0 \phi - A_0 \partial_1 \phi \Big).
\end{equation}
After integration by parts, and keeping the boundary terms, we obtain
\begin{equation}
    S_{\text{CS}} = \kappa \int dr dt \ \phi F_{10} + \frac{\kappa}{2} \int dr \big[ \phi A_1 \big]\bigg|^{t=\infty}_{t=-\infty} + \frac{\kappa}{2} \int dt \big[ \phi A_0 \big]\bigg|^{r=R}_{r=0},
\end{equation}
where $F_{10} = \partial_r A_0 - \partial_0 A_1$. Discarding the boundary terms, what we have is a 2-dimensional $BF$-theory, with $\phi$ acting as an on-shell constant Lagrange multiplier imposing the flatness condition on the 2-dimensional field strength. Note that $\phi$ is dimensionless, so this reduced theory has no length scale. \\ 

We can also identify the end-points of the radial coordinate, in which case we get an interesting situation. We relabel $x_1 = r \sim x_1+ 2\pi R$, and $x_2 = t$ to write the reduced action as
\begin{equation}
    S_{\text{CS}} = \kappa \int_{S^1 \times \mathbb R} d^2x \phi F_{12} = 2\kappa \int_{S^1 \times \mathbb R} \phi F.
\end{equation}
The second boundary term disappears because upon compactifying $r$, we identify the ansatz functions at $r=0$ and $r=R$. The first boundary term is not written above so the full action we are interested in reads
\begin{equation}
    S_{\text{CS}} = 2\kappa \int_{S^1 \times \mathbb R} \phi F + \frac{\kappa}{2} \int_0^R dr [\phi A_1] \bigg|_{t=-\infty}^{t=\infty}
\end{equation}
To see why this is interesting, let us obtain the first term of $S_{\text{CS}}$ in a familiar way. Under a gauge transformation $A \xrightarrow{} A + d \chi$, the boundary term coming from (\ref{U(1) CS action on X}) reads as
\begin{equation}
    \delta I = \frac{\kappa}{4\pi} \int_X  d\chi \wedge F = \frac{\kappa}{4\pi} \int_{\partial X} \chi F = \frac{\kappa}{4\pi} \int_{\mathbb R \times S^1} \chi F.
\end{equation}
The action we obtained from the symmetry reduction is similar to the boundary term of the Abelian Chern-Simons theory under a gauge transformation. The boundary term is known to be related to rational conformal field theory (RCFT) due to the edge modes \cite{Kapustin:2010if, Witten:1988hf, Elitzur-Shmuel-Moore-Schwimmer-Seiberg:1989nr, Tong:2016kpv}. The similarity of the boundary action and the action obtained via the disk reduction is curious, which calls for a more detailed investigation.

\subsection{\texorpdfstring{$SU(2)$}{SU(2)} Chern-Simons Theory on a Disk} \label{subsec 6.2: Disk reduction of SU(2) CS}

In section \ref{subsec 2.1: S1 * S2 reduction}, we considered Chern-Simons theory on $S^1 \times S^2$. Instead of $S^2$, one might consider an arbitrary 2-manifold $\Sigma$ and define the Chern-Simons theory on $X = S^1 \times \Sigma$. The immediate choice would be $\Sigma = D_2$ with $D_2$ a disk of radius $R$. \\

For $SU(2)$ Chern-Simons theory, we start on $X= \mathbb R \times D_2$. Making a disk reduction corresponds to using the conformal equivalence $D_2 - \{ 0 \} \sim I_R \times S^1$, which we discussed in the beginning of this section. In the action, we fix the $A_0 = 0$ gauge hence we have:
\begin{equation}
    S = \frac{\kappa}{8\pi} \int dt \int_{D_2} d^2x \varepsilon^{ij}\ A_i^a \frac{d}{dt} A_j^a,
\end{equation}
where $x_i\ (i=1,2)$ denote the coordinates on the disk. Now we will make a symmetry reduction such that the gauge fields are independent of the angular coordinate on the disk:
\begin{equation} \label{Ansatz on A_i^b and A_i^3 for SU(2) CS on the disk}
\begin{aligned}
    A^b_i &= \varepsilon_{ib}\ \frac{1}{r}\ \phi_A(r,t) + \frac{\left( \delta_{ib} - 2 \frac{x_ix_b}{r^2} \right)}{r} \chi_A(r,t) + \frac{x_ix_b}{r^2} A_r(r,t) , \\
    A^3_i &= \varepsilon_{ij} \frac{x_j}{r^2} \phi_B(r,t) + \frac{x_i}{r} B_r(r,t),
\end{aligned}
\end{equation}
with $b = 1,2$; and $r^2 = x_1^2 + x_2^2$. We have two distinct connections and associated scalar fields $\phi_A,\chi_A, \phi_B$. The reduced action reads
\begin{equation}
    S = \frac{\kappa}{4\pi} \left[ \int_{ I_R \times \mathbb R} d^2x ( \phi_A \partial_t A_r - \partial_t \phi_A A_r) + \int_{I_R \times \mathbb R} d^2x ( \phi_B \partial_t B_r - \partial_t \phi_B B_r) \right].
\end{equation}
The function $\chi_A$ that appears in the ansatz does not appear in the Chern-Simons term. One might shift $A_r \xrightarrow{} A_r + \frac{\chi_A}{r}$ in the ansatz for $\chi_A$ to appear, but we leave it as it
is\footnote{Shifting $A_r \xrightarrow{} A_r + \frac{\chi_A}{r}$ corresponds to changing the middle term of $A_i^b$ in (\ref{Ansatz on A_i^b and A_i^3 for SU(2) CS on the disk}) as $(\delta_{ib} - 2 x_ix_b/r^2) \xrightarrow{} (\delta_{ib} - x_ix_b/r^2$). While this is perfectly valid, this version of the ansatz gives a reduced action that contains $r$ explicitly. We do not want explicit $r$ dependency so we make the ansatz as in (\ref{Ansatz on A_i^b and A_i^3 for SU(2) CS on the disk}).}.
Doing integration by parts and interpreting $A_r$, $B_r$ as the spatial components of the connections of two independent $U(1)$ fields, and treating $(F_A)_{t,r} \equiv \partial_t A_r$, $(F_B)_{t,r} \equiv \partial_t B_r$ as the curvatures in the gauge $A_t = 0$, $B_t = 0$, we can write the reduced action as
\begin{equation}
    S = \frac{\kappa}{4\pi} \left[ \int_{I_R \times \mathbb R} d^2x \Big( \phi_A (F_A)_{t,r} + \phi_B (F_B)_{t,r} \Big) - \int_{I_R} dr \Big( \phi_A A_r + \phi_B B_r \Big) \bigg|_{t=-\infty}^{t=\infty} \right].
\end{equation}
In the differential form language, this reads
\begin{equation}
    S = \frac{\kappa}{2\pi} \left[ \int_{I_R \times \mathbb R} \Big( \phi_A F_A + \phi_B F_B \Big) - \frac{1}{2} \int_{I_R} \Big( \phi_A A + \phi_B B \Big) \bigg|_{t=-\infty}^{t=\infty} \right].
\end{equation}
We may identify the end-points of the $t$-coordinate if we wish to do so, in which case the above action is the reduction of $SU(2)$ Chern-Simons on $S^1 \times D_2$ to a $U(1) \times U(1)$ $BF$-theory without boundary terms (boundary terms disappear when we identify the end-points of $t$). On the other hand, we may leave $t$ uncompactified, and identify the end-points of $I_R$ to obtain a circle of radius $R$ and get a $BF$-type theory on $S_R^1 \times \mathbb R$. \\

The form of the reduced action is intriguing, as it signals a deeper relation between Chern-Simons theories and Conformal Field Theories (CFTs), in the sense that not only the boundary terms of Chern-Simons theory give CFTs, but also the bulk Chern-Simons theory with a disk reduction gives an action that is closely related to a CFT. Based on the results of sections \ref{sec 2: Euclidean SU(2) CS Theory on S1 * S2} and \ref{sec 3: Symmetry Reduction at the Semiclassical Regime}, where we showed that the partition functions of the reduced theories agree with those of the original theories under certain circumstances, one may suspect similar results to hold for the disk reductions discussed in this section as well. It is not immediately obvious whether such an agreement continues to hold for disk reductions. Nevertheless, that this reduction scheme gives a $BF$-type theory, which is closely related to the boundary CFT, indicates a strong relation between symmetry reductions of Chern-Simons theory (and, perhaps, those of Topological Field Theories) to the interplay between boundary CFTs and bulk TFTs \cite{Kapustin:2010if, Witten:1988hf, Elitzur-Shmuel-Moore-Schwimmer-Seiberg:1989nr, Tong:2016kpv, Dijkgraaf:1989pz}. 

\section{Symmetry Reduction of Euclidean \texorpdfstring{$SU(2)$}{SU(2)} Chern-Simons-Higgs Theory} \label{sec 6: SU(2) Chern-Simons Higgs and their monopoles}
In Euclidean space, we consider the action
\begin{equation} \label{Chern-Simons Higgs action in 3d}
    S = \frac{i\kappa}{8\pi g_{\text{YM}}^2} \int d^3x \varepsilon^{ijk} \left( A_i^a \partial_j A_k^a + \frac{\varepsilon^{abc}}{3} A_i^a A_j^b A_k^c \right) + \frac{1}{g_{\text{YM}}^2} \int d^3x  \left(  \frac{1}{2} D_i \Phi^a D_i \Phi^a + V(\Phi^a)  \right),
\end{equation}
where $\Phi^a$ is in the adjoint representation of $SU(2)$
\begin{equation}
    D_i \Phi^a = \partial_i \Phi^a + \varepsilon^{abc} A_i^b \Phi^c.
\end{equation}
We take the potential as the super-renormalizable Higgs potential
\footnote{We will take the renormalizable $(\Phi^a\Phi^a)^3$ term into consideration in subsection \ref{subsec 6.1: CS-Higgs on M = R * S2}. }
\begin{equation}
    V(\Phi^a) = \frac{\lambda}{4} ( \Phi^a \Phi^a - v^2)^2.
\end{equation}
The field equations obtained by varying with respect to the fields are
\begin{equation}
    \begin{aligned}
        0 &= \frac{\delta S}{\delta \Phi^a} \Longrightarrow 0= - D_i D_i \Phi^a + \frac{\partial V}{\partial \Phi^a}, \\
        0 &= \frac{\delta S}{\delta A_i^a} \Longrightarrow 0 = \frac{i\kappa}{4\pi g_{\text{YM}}^2} \varepsilon_{ijk} F_{jk}^a + \frac{1}{g_{\text{YM}}^2} \varepsilon^{abc} \Phi^b D_i \Phi^c.
    \end{aligned}
\end{equation}
Under a gauge transformation, the Higgs term is invariant. For the Chern-Simons term to be invariant mod $2\pi i m$ with $m$ an integer, we need to ensure
\begin{equation}
    k \equiv \frac{\kappa}{g_{\text{YM}}^2} \in \mathbb Z.
\end{equation}
Are there instanton solutions in the theory defined by (\ref{Chern-Simons Higgs action in 3d})? We know that if the kinetic term of the gauge sector was the 3d Yang-Mills term, then the corresponding Yang-Mills-Higgs action could be considered as the energy functional of a $3+1$ dimensional Georgi-Glashow model which admits monopoles that are called 't Hooft-Polyakov monopoles. On another route, one could take the 3d Yang-Mills Higgs action to be a $3+0$ dimensional Euclidean theory, where the solutions would be now 't Hooft-Polyakov instantons, strictly speaking. It is well known in the literature of topological solitons \cite{Manton:2004tk} that there are finite energy static monopole solutions in 3+1 dimensions (or finite Euclidean action instanton solutions in 3+0 dimensions), with a corresponding Bogomolny'i bound on the energy (the Euclidean action). In the limit where the Higgs potential vanishes, i.e. the BPS limit, one can saturate the bound if the fields solve the BPS equations 
\begin{equation}
    \frac{1}{2} \varepsilon_{ijk} F_{jk}^a = D_i \Phi^a,
\end{equation}
and these solutions are the BPS monopoles (instantons). The energy (Euclidean action) of the monopoles (instantons) for the BPS solutions is proportional to the topological charge of the soliton, and inversely proportional to the coupling constant $g_{\text{YM}}$, a characteristic scenario in topological solitons.  \\

Having given a brief review of what would have happened if the gauge sector had a Yang-Mills term, we now get back to our situation. The field equations of the Chern-Simons-Higgs theory from the variation of $A_i^a$ can be written as \cite{Edelstein:1994dy}
\begin{equation}
    B_i^a \Phi^a = 0,
\end{equation}
with $B_i^a = \frac{1}{2} \varepsilon_{ijk} F_{jk}^a$. This equation says that the magnetic field is orthogonal, in the sense of the quadratic form on the Lie algebra, to the Higgs field everywhere. If we consider 't Hooft's definition of the Abelian field strength in the direction of symmetry-breaking within the context of Chern-Simons-Higgs theory
\begin{equation}
    \mathcal F_{ij} = \mathcal F_{ij}^{(1)} + \mathcal F_{ij}^{(2)},
\end{equation}
with 
\begin{equation} \label{'t Hooft definition of abelian field strength}
    \begin{aligned}
        \mathcal F_{ij}^{(1)} &= \frac{\Phi^a}{(\Phi^b \Phi^b)^{1/2}} F_{ij}^a, \\
        \mathcal F_{ij}^{(2)} &= - \varepsilon^{abc} \frac{\Phi^a}{(\Phi^a \Phi^a)^{3/2}} D_i \Phi^b D_j \Phi^c,
    \end{aligned}
\end{equation}
we see that from the field equations we have $\mathcal F^{(1)} = 0$ for any solution. Since the magnetic flux of the monopole\footnote{In the context of Chern-Simons Higgs theory, we will abuse terminology and use monopoles and instantons interchangeably.}
is given by the surface integral of $\mathcal B_i = \frac{1}{2} \varepsilon_{ijk} \mathcal F_{jk}$ on a sphere at infinity, and since $\mathcal B_i^{(1)} \equiv \frac{1}{2} \varepsilon_{ijk} \mathcal F_{jk}^{(1)}$ is 0 everywhere, the only way to get a monopole with non-trivial magnetic flux is to make sure that $\mathcal B_i^{(2)} \equiv \frac{1}{2} \varepsilon_{ijk} \mathcal F_{jk}^{(2)}$ is non-vanishing at the boundary so that the surface integral of $\mathcal B = \mathcal B^{(1)} + \mathcal B^{(2)}$ is nonzero. But, if $\mathcal B^{(2)}$ is to be non-vanishing at infinity, then $D_i \Phi^a$ must be non-vanishing at infinity. Observe that the Euclidean action of the monopole solution goes like 
\begin{equation}
    S_E \sim \int d^3x (D_i \Phi^a)^2 + \cdots,
\end{equation}
at large $|x|$. But this action cannot converge if we want to construct a monopole solution with non-trivial flux at infinity, because for such a solution we necessarily have $D_i \Phi^a \xrightarrow{|x| = \infty} O(1/r)$, which means
\begin{equation}
    S_E \sim 4\pi \int dr r^2 O(1/r^2) + \cdots \xrightarrow{} \infty.
\end{equation}
So $S_E$ diverges linearly in $r$. Hence, it appears that the finiteness of the Euclidean action and the requirement of a non-trivial flux at infinity are in clash with each other. \\ 

Although we have shown that for Chern-Simons-Higgs theories one cannot have finite action monopoles with non-trivial flux, let us proceed in any case to construct some solutions with non-trivial flux. For a monopole-instanton type of solution, we employ the spherically symmetric ansatz that is well studied in the literature (see \cite{Tekin:1998qy, Pisarski:1986gr, Edelstein:1994dy} and the references therein)
\begin{equation}
\begin{aligned}
    A^a &= \Bigg[ \varepsilon_{iak} \frac{x_k}{r^2} \big( \varphi_2(r) - 1 \big) + \frac{\delta^{\perp}_{ia}}{r} \varphi_1(r)  + \frac{x_ix_a}{r^2} A(r) \Bigg] dx^i, \\
    \Phi^a &= \frac{x^a}{r} \Phi(r).
\end{aligned}
\end{equation}
The action reduces, upon defining $Y = \varphi_1+ i\varphi_2$, to:
\begin{equation}
    S = \frac{\kappa}{g_{\text{YM}}^2} \int dr \Big( \overline Y (\partial_r + iA) Y - i A \Big) + \frac{4\pi}{g_{\text{YM}}^2} \int dr \Big( \frac{r^2}{2} \Phi'^2 + \Phi^2 \overline Y Y \Big) + \int dr V.
\end{equation}
with
\begin{equation}
    \int dr V = \frac{\pi \lambda}{g_{\text{YM}}^2} \int dr r^2  (\Phi^2- v^2)^2 .
\end{equation}
Because the kinetic term of $Y$ is that of a fermionic field, we associate with the complex scalars $Y$ the Grassmannian fields $\psi$ and write
\footnote{At the level of the classical equations, this is no problem: One can pair the classical solutions of the two actions in a well-defined one-to-one manner.}
\begin{equation} \label{Reduced action of CSH mapped to the action with fermions}
    S =  \frac{1}{g_{\text{YM}}^2} \int dr \left( \frac{1}{2} 4\pi r^2 \Phi'^2 + \kappa \overline \psi (\partial_r + iA) \psi + 4\pi \Phi^2 \overline \psi \psi - i\kappa A \right) + \int dr V.
\end{equation}
If it were not for the $r^2$ factor in front of the $\Phi'^2$ term, this would be the action of a single scalar field in 1-dimension coupled to a Grassmannian valued field, and the $U(1)$ symmetry of the fermionic action is gauged with a Chern-Simons kinetic term. Since $\Phi$ is real, it cannot couple to the $U(1)$ gauge field $A$. With the identification $\tau = r$, and scaling the fields $\Psi(\tau) = (\sqrt\kappa/g_{\text{YM}}) \psi(\tau)$, and $\mathsf X(\tau) = (2\sqrt{\pi}/g_{\text{YM}}) \Phi(\tau)$ we can rewrite (\ref{Reduced action of CSH mapped to the action with fermions}) as
\begin{equation} \label{Reduced action of CSH with the scaled fields}
    S = \int d\tau \left( \frac{\tau^2}{2} \dot{\mathsf{X}}^2 + \overline{\Psi} (\partial_\tau + i A) \Psi - \frac{i\kappa}{g_{\text{YM}}^2} A + \frac{g_{\text{YM}}^2}{\kappa} \mathsf{X}^2 \overline \Psi \Psi \right) + \int d\tau V(\mathsf X),
\end{equation}
with a 1-dimensional $r$-dependent Higgs potential of which the minima are shifted
\begin{equation}
    \int d\tau V(\mathsf X) = \frac{\lambda g_{\text{YM}}^2}{16\pi} \int d\tau \tau^2 \left( \mathsf X^2 - \frac{4\pi v^2}{g_{\text{YM}}^2} \right)^2.
\end{equation}
Based on the interaction between $\mathsf X$ and $\Psi$, we make the following observation about the reduced action (\ref{Reduced action of CSH with the scaled fields}). If one were to add a potential of the form $\frac{1}{2} h'(\mathsf X)^2 \sim \mathsf X^6$, the interaction terms between $\mathsf X$ and $\Psi$ can be written as $ \frac{1}{2} h'(\mathsf X)^2 + h''(\mathsf X) \overline \Psi \Psi$ with $h$ being the superpotential \cite{Hori:2003ic}. With the canonical kinetic terms of the bosonic and fermionic fields, this gives a supersymmetric quantum mechanical model in the Euclidean formalism. In the next section, we will cure the explicit $r$-dependencies and recast the symmetry-reduced Chern-Simons-Higgs action as the action of a supersymmetric quantum mechanical model. 

\subsection{Chern-Simons-Higgs Theory on \texorpdfstring{$\mathbb R^+ \times S^2$}{R+ * S**2}}  \label{subsec 6.1: CS-Higgs on M = R * S2}

To remove the $r$-dependence in the kinetic term of $\Phi$ in (\ref{Reduced action of CSH mapped to the action with fermions}), we define the theory on $\mathcal M = \mathbb R^+ \times S^2$. To see the motivation for this choice, we recall the conformal equivalence discussed in section \ref{sec 2: Euclidean SU(2) CS Theory on S1 * S2} that is $\mathbb R^3-\{0\} \sim \mathbb R^+ \times S^2$. While doing a spherically symmetric reduction, we are using this equivalence. The fields have components along $\mathbb R^+$ and along $S^2$, and by imposing them to be independent of $S^2$, we get a reduced action on $\mathbb R^+$. When we impose the gauge potential to be independent of $S^2$, its components along the sphere become scalar fields $\varphi_1,\varphi_2$ which we combine into a complex scalar field $Y = \varphi_1+ i\varphi_2$. The gauge symmetry is broken from $SU(2)$ to $U(1)$, and the latter acts on $Y,\overline Y$ as usual.  \\ 

We thus consider the action
\begin{equation} \label{Chern-Simons Higgs on M = R * S2}
\begin{aligned}
    S &= \frac{i\kappa}{8\pi g_{\text{YM}}^2} \int_{\mathcal M} d^3x \varepsilon^{ijk} \left( A_i^a \partial_j A_k^a + \frac{\varepsilon^{abc}}{3} A_i^a A_j^b A_k^c \right) \\
    &\hspace{4mm} + \frac{1}{ g_{\text{YM}}^2 } \int_{\mathcal M} d^3x \sqrt{|g|} \left( \frac{1}{2} g^{ij} D_i \Phi^a D_j \Phi^a + V(\Phi)  \right).
\end{aligned}
\end{equation}
Now the metric of $\mathcal M$ appears in the Higgs term. To compute the reduced action, we pass to the dreibein formalism, and exploit the similarity in the index structure between $A_i^a$ and $e_i^a$. We thus write an ansatz for $e$:
\begin{equation}
    e^a = \Bigg[ -\varepsilon^a_{\hspace{2mm}ik} \frac{x^k}{r^2} e_1(r) +  \frac{\delta^{\perp a}_i}{r} e_2(r) + \frac{x_ix^a}{r^2} e_3(r) \Bigg] dx^i.
\end{equation}
Observe that $x^i e_i^a = e_3 x^a$ based on the ansatz above. We want $x^a = x^i e_i^a$ so that $\delta^{ab} = e_i^a e_j^b g^{ij}$ hence we fix $e_3 = 1$. We find the metric corresponding to this dreibein as ($r>0$)
\begin{equation}
\begin{aligned}
    ds^2 = dr^2 + (e_1^2+e_2^2) d\Omega_2.
\end{aligned}
\end{equation}
Hence, to get the metric of $\mathbb R^+ \times S^2$, we fix $e_1^2+e_2^2 = r_0^2$, with $r_0$ being the radius of $S^2$. We insert the ans\"atze on the fields and the dreibein (\ref{Chern-Simons Higgs on M = R * S2}), to get 
\begin{equation} \label{Reduced CSH action on M = R * S2 in terms of the complex scalar Y}
\begin{aligned}
    S = \frac{\kappa}{g_{\text{YM}}^2} \int dr \Big( \overline Y (\partial_r + iA) Y - i A \Big) + \frac{4\pi}{g_{\text{YM}}^2} \int dr \left( \frac{r_0^2}{2} \Phi'^2 + \Phi^2 \overline Y Y + V \right),
\end{aligned}
\end{equation}
with the reduced potential $V = \frac{\lambda r_0^2}{4} (\Phi^2 -v^2)^2$. Observe that we have cured the explicit $r$-dependencies from the kinetic term of $\Phi$ and $V$, through the choice of the geometry over which the theory is defined. Indeed, the flat space limit corresponds to $(e_1^2+e_2^2) = r^2$ so we recover (\ref{Reduced action of CSH with the scaled fields}). It makes sense that the choice $\mathbb R^+\times S^2$ of geometry removes the $r^2$ term in front of $\Phi'^2$ and the potential $V$ because that factor resulted from the angular integral which gives $4\pi r^2$; but if we fix the geometry such that all spheres have the same surface area, $4\pi r_0^2$, then the reduced action comes with $r_0^2$, not $r^2$. If we add the Einstein-Hilbert term into the action, making the geometry dynamical, then we would also need a spin connection $\omega$ which we could also write in the form of $e_i^a$. We could reduce the Einstein-Hilbert action in the first-order formalism to a 1d theory to attempt to solve the equations of the Einstein-Hilbert-Chern-Simons-Higgs theory. In \cite{Ferstl:2000qp}, this was studied with the gauge sector having the Yang-Mills term as well. We will not take gravity to be dynamical here. \\

We can compactify $\mathbb R^+$ to a circle of radius $\beta$. When the radial component is compactified, the Euclidean action of the monopoles will not be infinite but will be proportional to $\beta$, with the constant of proportionality related to the string tension of a monopole-antimonopole pair \cite{Edelstein:1994dy}. Identifying the end-points of $r$, we get periodic boundary conditions on the fields of the reduced theory.\\

Associating $Y$ with a Grassmannian field $\psi$ one gets
\begin{equation}
    S = \frac{\kappa}{g_{\text{YM}}^2} \int dr \bigg( \overline \psi (\partial_r + iA) \psi - i A  \bigg) + \frac{4\pi}{g_{\text{YM}}^2} \int dr \left( \frac{r_0^2}{2} \Phi'^2 + \Phi^2 \overline \psi \psi + \frac{\lambda r_0^2}{4} (\Phi^2 -v^2)^2 \right),
\end{equation}
and scaling the fields by $\Psi = \frac{\sqrt \kappa}{g_{\text{YM}}} \psi$, $\mathsf X = \frac{2\sqrt\pi r_0}{g_{\text{YM}}} \Phi$,
\begin{equation} \label{1d model with psi, X, and interaction 1/r_0k}
\begin{aligned}
    S &= \int d\tau \left( \overline \Psi (\partial_\tau + i A) \Psi - \frac{i\kappa}{g_{\text{YM}}^2} A \right) \\
    & \hspace{2mm} + \int d\tau \left( \frac{1}{2} \dot{\mathsf X}^2 + \frac{g_{\text{YM}}^2}{r_0^2\kappa} \mathsf X^2 \overline \Psi \Psi + \frac{g_{\text{YM}}^2 \lambda }{16\pi r_0^2} \left(  \mathsf X^2 - \frac{4\pi r_0^2 v^2}{g_{\text{YM}}^2}\right)^2 \right).
\end{aligned}
\end{equation}
The expectation value for the Higgs field of the 3-dimensional theory is $v$ whereas that of the $1$-dimensional reduced theory is $4\pi r_0^2/g_{\text{YM}}^2 v$, so it is scaled. We would like to recast this action as the action of a supersymmetric quantum mechanical model, based on our observation earlier in this section. To have supersymmetry, we need to add a $\mathsf X^6$ term into the potential, which corresponds, in the Chern-Simons Higgs theory, to add a $(\Phi^a \Phi^a)^3$ term to the potential. We will fix the coefficient of this potential so that the reduced action has a supersymmetry. The interaction between $\Psi$ and $\mathsf X$ implies the following superpotential
\begin{equation} \label{superpotential of reduced CSH theory}
    h(\mathsf X) = \frac{1}{12r_0^2 k} \mathsf X^4.
\end{equation}
To include $\frac{1}{2} (h')^2$ in the reduced action, we add a sixth order potential $V = \frac{8\pi^2}{9\kappa^2} (\Phi^a \Phi^a)^3$ to the double well potential, so the full potential of the original theory reads 
\begin{equation}
    V = \frac{\lambda}{4} (\Phi^a \Phi^a - v^2)^2 + K (\Phi^a \Phi^a)^3, \quad K = \frac{8\pi^2}{9\kappa^2}.
\end{equation}
Then, in the formal limit $\lambda \xrightarrow{} 0$, the reduced theory could be identified with a supersymmetric quantum mechanical model having a superpotential (\ref{superpotential of reduced CSH theory}), in the sense that the solutions of the reduced theory can be mapped to the solutions of the supersymmetric quantum mechanics. \\

Actually, the reduced theory does not have a supersymmetry in its present form, because of the $A \overline \Psi \Psi$ term. We can eliminate $A$ from the action (\ref{1d model with psi, X, and interaction 1/r_0k}), which twists the boundary condition of $Y$ and $\Psi$: $\Psi(\beta) = e^{-i\int_0^\beta dr A(r)} \Psi(0)$. Eliminating $A$, and taking the formal limit $\lambda \xrightarrow{} 0 $ so that the double well term vanishes, the reduced action of the Chern-Simons-Higgs theory can be identified with
\begin{equation}
    S_{E} = \int d\tau \left( \frac{1}{2} \left(\frac{d\mathsf X}{d\tau} \right)^2 + \overline \Psi \frac{d}{d\tau} \Psi + \frac{1}{2} h'^2(\mathsf X) + h''(\mathsf X) \overline\Psi \Psi \right).
\end{equation}
This action is invariant under the supersymmetry transformations
\begin{equation}
\begin{aligned}
    \delta \mathsf X = \epsilon \overline \Psi - \overline \epsilon \Psi, \quad 
    \delta \Psi = \epsilon \left( -\frac{d\mathsf X}{d\tau} + h'(\mathsf X) \right), \quad
    \delta \overline \Psi = \overline \epsilon \left( + \frac{d \mathsf X}{d\tau} + h'(\mathsf X) \right),
\end{aligned}
\end{equation}
and we have the superpotential $h(\mathsf X) = \frac{1}{12r_0^2k} \mathsf X^4$. Hence, the Lagrangian reads
\begin{equation} \label{Reduced susyqm action with h = X^4}
    L = \frac{1}{2} \left( \frac{d \mathsf X}{d\tau} \right)^2 + \overline \Psi \frac{d}{d\tau} \Psi + \frac{1}{r_0^2 k} \mathsf X^2 \overline\Psi\Psi + \frac{1}{18 r_0^4 k^2} \mathsf X^6 .
\end{equation}
The action (\ref{Reduced CSH action on M = R * S2 in terms of the complex scalar Y}) governs the spherically symmetric monopole solutions in the theory defined by the action (\ref{Chern-Simons Higgs on M = R * S2}), and the solutions of (\ref{Reduced CSH action on M = R * S2 in terms of the complex scalar Y}) can be identified with the solutions of (\ref{Reduced susyqm action with h = X^4}). Thus, the Chern-Simons Higgs instanton equations on $\mathcal M = S_\beta^1 \times S^2$ can formally be viewed as the equations of a supersymmetric quantum mechanical model. \\

The addition of a sixth-order term in the potential does not change the fact that $\mathcal F_{ij}^{(1)}$ in (\ref{'t Hooft definition of abelian field strength}) is 0 for all solutions to the field equations, nor does it change the expression for $\mathcal F_{ij}^{(2)}$. So, inserting our ansatz into (\ref{'t Hooft definition of abelian field strength}), we find the magnetic field associated to $\mathcal F^{(2)}$ as
\begin{equation}
    \mathcal B_i^{(2)} = - \frac{x_i}{r^3} \overline Y Y,
\end{equation}
the flux of which is 
\begin{equation}
    \Phi_{\text{flux}} = -4\pi(\varphi_1^2+\varphi_2^2) = -4\pi \overline Y Y.
\end{equation}
So, $\overline Y Y$ is associated with the magnetic charge of the instanton solution. In terms of the fermionic field $\Psi$, we have
\begin{equation}
    \Phi_{\text{flux}} = - \frac{4\pi}{k} \overline \Psi \Psi.
\end{equation}
Because the identification between the Chern-Simons Higgs instanton equations, and the equations of the supersymmetric quantum mechanics hold only at the classical level, this is only an interesting observation about the form of Chern-Simons Higgs instanton equations and is unlikely to survive at the quantum level.

\subsection{A new perspective on the reduced Chern-Simons-Higgs Theory} \label{subsec 6.2: A new perspective on the reduced CSH theory}
The similarity between the instanton equations and the classical equations of a supersymmetric model that we established in the previous subsection is valid only at the level of classical equations. Can we play with the partition function based on the action (\ref{Reduced CSH action on M = R * S2 in terms of the complex scalar Y}) to obtain a fermionic theory at the quantum level as we did in subsection \ref{subsec 2.2: 1d theory path integral as a susy qm index}? \\

We will consider the partition function of the symmetry reduced theory from $\mathbb R^+ \times S^2$ to $S^1_\beta$, where we compactify $\mathbb R^+$, so we have the periodic boundary conditions $Y(\beta) = Y(0)$. We will eliminate $A$ from the action, which gives the twisted boundary conditions $Y(\beta) = e^{-i\int_0^\beta dr A(r)} Y(0)$. So the partition function we consider reads
\begin{equation}
    Z_{\text{CSH}}(\mathbb R^+\times S^2 \xrightarrow{} S_\beta^1) = \int \mathcal D \Phi \int  \mathcal D Y \mathcal D \overline Y\bigg|_{Y(\beta) = e^{-i\int_0^\beta dr A(r)} Y(0) } e^{-S},
\end{equation}
with the Lagrangian
\begin{equation}
\begin{aligned}
    L &=  \overline Y \mathsf D Y + \mathcal L(\Phi), \\
    \mathcal L(\Phi) &= \frac{2\pi r_0^2}{g_{\text{YM}}^2 } \Phi'^2 + V(\Phi),
\end{aligned}
\end{equation}
where we defined the differential operator $\mathsf D = \left( k \partial_r + \frac{4\pi}{g_{\text{YM}}^2} \Phi \right)$, with $k = \frac{\kappa}{g_{\text{YM}}^2}$ being the Chern-Simons level, and we leave the potential $V(\Phi)$ unspecified for the moment. Performing the $Y$ integrations, we get
\begin{equation}
    Z_{\text{CSH}}(\mathbb R^+\times S^2 \xrightarrow{} S_\beta^1) = \int \mathcal D \Phi e^{- \int dr \mathcal L(\Phi)} \frac{1}{\det \mathsf D}.
\end{equation}
We multiply and divide by $\det (\mathsf D)$ to get $\frac{\det (\mathsf D)}{\det (\mathsf D^2)}$ and introduce new fields $\psi$,$\Upsilon$ to expand this ratio as a functional integral and obtain a partition function based on a new Lagrangian 
\begin{equation}
    Z_{\text{CSH}}(\mathbb R^+ \times S^2 \xrightarrow{} \mathbb R^+) = \int \mathcal D \Phi \mathcal D \Upsilon \mathcal D \overline \Upsilon \mathcal D \psi \mathcal D \overline \psi \ e^{- \mathcal S},
\end{equation}
where
\begin{equation}
    \mathcal S = \int dr \left[ -\overline{\mathsf D \Upsilon} \mathsf D\Upsilon + \overline \psi \mathsf D \psi + \frac{2\pi r_0^2}{g_{\text{YM}}^2 } \Phi'^2 + V(\Phi) \right].
\end{equation}
We assume $k>0$ and rescale the fields by $\Psi = \sqrt k \psi$, $\mathsf X = \frac{2\sqrt{\pi} r_0}{g_{\text{YM}}} \Phi$, and identify $\tau = r$ to get
\begin{equation}
    \mathcal S = \int d\tau \ \left( -k^2 \left| \left( \frac{d}{d\tau} + s \mathsf X \right) \Upsilon \right|^2 + \frac{1}{2} \left( \frac{d\mathsf X}{d\tau} \right)^2 + \overline \Psi \frac{d}{d\tau} \Psi - s \mathsf X \overline \Psi \Psi + V \right),
\end{equation}
where we defined $s = \frac{2\sqrt{\pi}}{g_{\text{YM}}^2 r_0 k}$. If we choose $V$ such that $V = \frac{s^2}{8} \mathsf X^4$, we can write this action as 
\begin{equation} \label{cal S = Upsilon + X + Psi with (X,Psi) having almost susy}
    \mathcal S = \int d\tau  \left( -k^2 \left| \left( \frac{d}{d\tau} + s \mathsf X \right) \Upsilon \right|^2 + \frac{1}{2} \left( \frac{d\mathsf X}{d\tau} \right)^2 + \overline \Psi \frac{d}{d\tau} \Psi + \frac{d^2 h}{d\mathsf X^2} \ \overline \Psi \Psi + \frac{1}{2} \left( \frac{d h}{d\mathsf X} \right)^2 \right),
\end{equation}
where $h = -\frac{s}{6} \mathsf X^3$. This action is the same as (\ref{Reduced CSH action on M = R * S2 in terms of the complex scalar Y}), at the level of the partition function, so this is a quantum mechanical duality between the two theories. Although one might be tempted to conclude that this is a supersymmetric action, the $\mathsf X$ term that appears in the kinetic term of the complex field $\Upsilon$ spoils this supersymmetry. One may try to give a variation $\delta \Upsilon$ under supersymmetry such that the term $|\mathsf D \Upsilon|^2$ is supersymmetry invariant, but what type of variation would leave such a term invariant if at all, is not immediately obvious. In any case, we have been able to show that the reduced action (\ref{Reduced CSH action on M = R * S2 in terms of the complex scalar Y}) has a fermionic character not only at the classical level but also at the quantum level.

\section{Symmetry Reduction of \texorpdfstring{$SL(2,\mathbb C)$}{SL(2,C)} Chern-Simons-Higgs Theory} \label{sec 7: SL(2,C) Chern-Simons Higgs and their monopoles}
Let us consider what happens when we reduce the $SL(2,\mathbb{C})$ Chern-Simons Higgs theory through our ansatz. We consider 
\begin{equation}
\begin{aligned}
    S = &\frac{t}{8\pi g_{\text{YM}}^2} \int_{\mathcal M} d^3x \varepsilon^{ijk} \left( A_i^a \partial_j A_k^a + \frac{\varepsilon^{abc}}{3} A_i^a A_j^b A_k^c \right) \\
    & \hspace{20mm}+ \frac{1}{g_{\text{YM}}^2 }\int_{\mathcal M} d^3x \sqrt{|g|} \left( \frac{1}{2} g^{ij}  D_i \Phi^a  D_j \Phi^a + V(\Phi) \right) \\ 
    + &\frac{\overline t}{8\pi g_{\text{YM}}^2} \int_{\mathcal M} d^3x \varepsilon^{ijk} \left( \overline A_i^a \partial_j \overline A_k^a + \frac{\varepsilon^{abc}}{3} \overline A_i^a \ \overline A_j^b \ \overline A_k^c \right) \\
    &\hspace{20mm} + \frac{1}{g_{\text{YM}}^2} \int_{\mathcal M} d^3x \sqrt{|g|} \left( \frac{1}{2} g^{ij} \overline{D_i  \Phi}^a \ \overline{D_j \Phi}^a + \overline V(\overline \Phi) \right),
\end{aligned}
\end{equation}
where the connection is a doublet of the form $(A, \overline A) = ( A^a T^a, \overline{A}^a T^a)$ and similarly for the adjoint field $(\Phi,\overline\Phi) = (\Phi^a T^a, \overline \Phi^a T^a)$. The Higgs field is in the adjoint representation, hence
\begin{equation}
\begin{aligned}
    D_i \Phi^a &= \partial_i \Phi^a + \varepsilon^{abc} \ A_i^b \ \Phi^c, \\
    \overline{D_i \Phi}^a &= \partial_i \overline \Phi^a + \varepsilon^{abc} \ \overline{A}_i^b \ \overline \Phi^c. 
\end{aligned}
\end{equation}
The same story unfolds with the following ans\"atze:
\begin{equation}
    \begin{aligned}
    A^a &=  \Bigg[ (\varphi_2 - 1) \varepsilon_{i}^{\hspace{2mm} ak} \frac{x_k}{r^2}  + \varphi_1 \frac{\delta^{\perp a}_{i}}{r} +  A \frac{x_ix^a}{r^2} \Bigg] dx^i, \\
    \overline{A}^a &=  \Bigg[ (\overline\varphi_2 - 1) \varepsilon_{i}^{\hspace{2mm} ak} \frac{x_k}{r^2}  + \overline\varphi_1 \frac{\delta^{\perp a}_{i}}{r} -  \overline A \frac{x_ix^a}{r^2} \Bigg] dx^i, \\
    \Phi^a &= \frac{x^a}{r} \Phi, \\
    \overline \Phi^a &= \frac{x^a}{r} \overline \Phi, \\
    e^a &= \Bigg[ -\varepsilon^a_{\hspace{2mm}ik} \frac{x^k}{r^2} e_1 + \delta^{\perp a}_i \frac{e_2}{r} + \frac{x_ix^a}{r^2} e_3 \Bigg] dx^i,
    \end{aligned}
\end{equation}
where all of the ansatz fields depend only on $r = (x_ix_i)^{1/2}$. We take $t = i\kappa$ and $\overline t = -i\kappa$, and invariance under large gauge transformations enforces the parameter $k \equiv \frac{\kappa}{g_{\text{YM}}^2}$ to be an integer. Choosing $\mathcal M = \mathbb R^+ \times S^2$, the action reduces to
\begin{equation} \label{Reduced action of SL(2,C) Chern-Simons Higgs with only scalars}
\begin{aligned}
    S &= \frac{\kappa}{g_{\text{YM}}^2} \int dr \Big( \overline Y_1 (\partial_r + iA) Y_1 - iA \Big) \\
    &+ \frac{4\pi}{g_{\text{YM}}^2} \int dr \left( \frac{r_0^2}{2} \Phi'^2 + \Phi^2 \overline Y_1 Y_1 + \frac{\lambda r_0^2}{4} (\Phi^2 - v^2)^2 \right) \\
    & + \frac{i\kappa}{g_{\text{YM}}^2} \int dr \Big( \overline Y_2 (\partial_r + i\overline A) Y_2 - i \overline A \Big) \\
    &  + \frac{4\pi}{g_{\text{YM}}^2} \int dr \left( \frac{r_0^2}{2} \overline \Phi'^2 + \overline \Phi^2 \overline Y_2 Y_2 + \frac{\lambda r_0^2}{4} (\overline \Phi^2 - v^2)^2\right).
    \end{aligned}
\end{equation}
At the classical level, we can again identify the equations following from this action to the equations that follow from the variation of a supersymmetric quantum mechanical action, this time with two superpotentials
\begin{equation}
\begin{aligned}
    h_1(\mathsf X_1) = \frac{1}{12 r_0^2 k} \mathsf X_1^4, \\
    h_2(\mathsf X_2) = \frac{1}{12 r_0^2 k} \mathsf X_2^4.
\end{aligned}
\end{equation}
This is provided if we fix the potentials to
\begin{equation}
\begin{aligned}
    V(\Phi) = \frac{\lambda}{4} (\Phi^a \Phi^a - v^2)^2 + \frac{8\pi^2}{9\kappa^2} (\Phi^a \Phi^a)^3, \\
    \overline V (\overline \Phi) =  \frac{\lambda}{4} (\overline \Phi^a \overline \Phi^a - v^2)^2 + \frac{8\pi^2}{9\kappa^2} (\overline \Phi^a \overline \Phi^a)^3 ,
\end{aligned}
\end{equation}
and take the formal limit $\lambda \xrightarrow{} 0$. We can also play with the path integral based on the action (\ref{Reduced action of SL(2,C) Chern-Simons Higgs with only scalars}) to write it as a theory with two real bosons, two complex bosons, and two complex fermions where the first set of complex bosons and the fermions obey the twisted boundary conditions with the twist $e^{-i\int_0^\beta dr A(r)}$ and the second set has the twist $e^{-i\int_0^\beta dr \overline A(r)}$, upon compactifying $\mathbb R^+$ to $S^1_\beta$ and eliminating $A,\overline A$ from the action. With this, we will get two independent sectors, both sectors having the action (\ref{cal S = Upsilon + X + Psi with (X,Psi) having almost susy}).

\section{Semiclassical Regime of Quantum Chern-Simons-Higgs Theory} \label{sec 8: Semiclassical limit of Quantum Chern-Simons Higgs theory}

By a semiclassical argument, we were able to show that the pure quantum Chern-Simons theory over $S^3$ in the large $\kappa$ limit agrees with the symmetry reduced theory; and we computed the partition function of the reduced theory from $S^1 \times S^2$ to be 1, in agreement with the original partition function for Chern-Simons theory on $S^1\times S^2$. Does such a correspondence work for the Chern-Simons-Higgs theory? It is considerably more involved to show what we have shown for the pure Chern-Simons theory. To sketch the level of involvement for carrying out the semiclassical evaluation of the Chern-Simons-Higgs theory, we argue based on a 0-dimensional QFT. Consider the following path integral
\begin{equation}
    Z = \int dx dy e^{-f(x,y)}.
\end{equation}
We denote the space of solutions $\delta f = 0$ as $\mathscr M$. For a pair $(x_n,y_n) \in \mathscr M$, one has
\begin{equation}
    f(x,y) = f(x_n,y_n) + \frac{1}{2} \partial_i \partial_j f(x_n,y_n) \delta x^i \delta x^j,
\end{equation}
with $i=(1,2)$ and $\delta x^i = x^i - x_n^i$. With this, the semiclassical path integral reads
\begin{equation}
    Z \approx \sum_{(x_n,y_n)\in \mathscr M} e^{-f(x_n,y_n)} \int_{B(x_n,y_n)} d[\delta x^1] d[\delta x^2] e^{-\frac{1}{2} \partial_i \partial_j f(x_n,y_n) \delta x^i \delta x^j}.
\end{equation}
To perform the integral over $\delta x^1$ and $\delta x^2$, we need to diagonalize the Hessian 
\begin{equation}
    H_{ij} = \partial_i \partial_j f(x_n,y_n).
\end{equation}
For this, one must ensure that $\det H \neq 0$. Even in that case, it is highly non-trivial to generalize this to the 3-dimensional Chern-Simons-Higgs theory. For one thing, in the $0$-dimensional theory, the fields $x,y$ do not have extra structure on them, they are just variables. On the other hand, for a QFT in 3-dimensions, the fields can have index structures allowed by the dynamical and internal symmetries that are proposed to be present in the theory. For example, in the 3d Chern-Simons-Higgs theory, one has a Lie-algebra valued 1-form gauge potential $A$ and an adjoint valued scalar field $\Phi$. It is not clear how one can combine these fields with different index structures when attempting to diagonalize the functional Hessian 
\begin{equation}
    \mathcal H = \frac{\delta^2 S_{\text{CS-Higgs}}}{\delta {\boldsymbol{\mathcal A}} \delta {\boldsymbol{\mathcal A}} }
\end{equation}
with $\boldsymbol{\mathcal A}$ representing the gauge and the Higgs fields. \\

But even before these discussions, one can make the following observation: In demonstrating that the $S^3$ Chern-Simons is equivalent to its 1d reduced theory, an essential part of the argument was the group structure in the space of critical points of the Chern-Simons action. This group structure is the crucial ingredient that allowed the results of section \ref{sec 3: Symmetry Reduction at the Semiclassical Regime} as we discuss there. Although we cannot say whether there is an exact agreement between the quantum Chern-Simons-Higgs and the symmetry-reduced theory, we can sketch an argument to show that they at least have an approximate relationship with each other. We argue as follows: We start with (\ref{Chern-Simons Higgs on M = R * S2}) and symmetry reduce to (\ref{Reduced CSH action on M = R * S2 in terms of the complex scalar Y}). Some subset of the solutions of (\ref{Chern-Simons Higgs on M = R * S2}) can thus be recast as the solutions of (\ref{Reduced CSH action on M = R * S2 in terms of the complex scalar Y}). Let us approximate the path integral of (\ref{Chern-Simons Higgs on M = R * S2}) by summing over only the classical solutions that possess spherical symmetry 
\begin{equation}
\begin{aligned}
    Z_{\text{CSH}} &= \int \mathscr D A \mathscr D \Phi e^{-S_{\text{CSH}}[\Phi, A]} \\
    &\approx \sum_{(\Phi_{\text{cl}}, A_{{\text{cl}}}) \in \mathscr M_{\text{CSH}}^r } e^{-S_{\text{CSH}}[\Phi_{\text{cl}},A_{{\text{cl}}}]} \det \Big( S_{\text{CSH}}''[\Phi_{\text{cl}}, A_{{\text{cl}}}] \Big)^{-1/2}.
\end{aligned}
\end{equation}
The determinant of the Hessian of the action and the sum over $\Phi_{\text{cl}}, A_{\text{cl}}$ are symbolic, where $\Phi_{\text{cl}}$ and $A_{\text{cl}}$ are classical solutions of the action $S_{\text{CS-Higgs}}$, and $\mathscr M_{\text{CS-Higgs}}^r$ denotes the solution space of the Chern-Simons Higgs theory with spherical symmetry. This is a weaker approximation than the one if we were to add over $\mathscr M_{\text{CS-Higgs}}$, the entire solution space. Since we are considering only the spherically symmetric solutions, the symbolic sum will give the semiclassical approximation of the quantum theory based on the action (\ref{Reduced CSH action on M = R * S2 in terms of the complex scalar Y}). Therefore, we write 
\begin{equation}
    Z_{\text{CS-Higgs}} \sim Z_{\text{reduced}},
\end{equation}
where $Z_{\text{reduced}}$ is the partition function of the reduced theory, and by $\sim$ we mean that the right-hand-side manages to capture only some portion of the left-hand-side.

\section{Conclusions and Discussions}
Let us give an overview of our results and discuss their implications. In the first part of the paper, we focused on pure Chern-Simons gauge theories defined on the topology of either $S^1 \times S^2$ or $S^3$. First, we have shown that the symmetry reductions of Chern-Simons theories with gauge group $SU(2)$ yield 1-dimensional quantum field theories with a $U(1)$ gauge group, but with a fermionic kinetic term for a bosonic field. We have shown in \ref{subsec 2.1: S1 * S2 reduction} that $Z(S^1 \times S^2) = 1 = Z(S^1 \times S^2 \xrightarrow{} S^1)$ where $Z(S^1 \times S^2)$ is the partition function of the original theory, and $Z(S^1 \times S^2 \xrightarrow{} S^1)$ is that of the reduced theory. This is an indication that the reduced theory captures the original one at the level of path integral. We do not have a result relating the observables of the two theories for $S^1 \times S^2$ (as opposed to the results for $S^3$). In \ref{subsec 2.2: 1d theory path integral as a susy qm index}, we show that the reduced action can be viewed as the action of a bosonic field and a fermionic field coupled to the same $U(1)$ field with the same charges, both obeying periodic boundary conditions. Moreover, we have given an interpretation of this path integral as a Witten index of a supersymmetric quantum mechanical model with a trivial superpotential. This interpretation is not useful computationally, but conceptually it has the potential to connect the path integral of Chern-Simons theory with a supersymmetric index. At this level, this is only a curious observation that should be explored. \\

For the $S^3$ Chern-Simons theory, we carried out the symmetry reduction and computed the partition function of the reduced theory. We then argued that the semiclassical limit of $Z(S^3)$ agrees with the semiclassical limit of $Z_{\text{spherical}}$, where $Z(S^3)$ is the original theory and $Z_{\text{spherical}}$ is the partition function where the fields are constrained to spherically symmetric configurations, which could effectively be described by a 1d reduced theory. In the semiclassical regime, we argued that the Wilson loop observables should match with some observables of the reduced theory, but what they correspond to in the reduced theory is an outstanding problem. Based on the results, we discuss possible generalizations for other gauge groups and other possible symmetry reductions in \ref{subsec 3.2: SYM reduction of G CS on S^3}. It turns out that the group structure in the solution space of Chern-Simons theory allows for semiclassical agreements between various theories. It is possible to hand-pick the space of flat connections based on a symmetry condition, restrict the partition function to only the field configurations that obey this symmetry, and if the gauge group is changed accordingly, one obtains a reduced QFT, whose observables in the semiclassical limit agree with the observables of $S^3$ Chern-Simons with some gauge group $G$. We discuss the role of the group structure in the space of critical points (before the modding out of gauge redundancy) of the Chern-Simons action, which is crucial. Based on the construction in section \ref{sec 3: Symmetry Reduction at the Semiclassical Regime}, it is not hard to see that quantum field theories based on an action for which the space of critical points of the action admits a group structure could allow similar reductions. An interesting question is then to find all QFTs for which the space of critical points of the action admits a group structure. For these theories, one can modify the argument constructed in section \ref{sec 3: Symmetry Reduction at the Semiclassical Regime} to show that such reduced QFTs would agree, at the semiclassical level, with the master theory from which one would make the symmetry reduction. \\

This correspondence between two quantum theories at the semiclassical limit is reminiscent of the AdS/CFT duality that has been very influential in understanding both the gauge and gravity theories in the past 25-30 years \cite{Maldacena:1997re, Witten:1998qj, Gubser:1998bc}. In the correspondence discussed here, we start with a 3-dimensional theory and reduce it to a lower-dimensional effective theory based on a symmetry condition, of which the observables in the semiclassical regime agree with those of the original theory. Our construction is valid for Chern-Simons theories on $S^3$. For different topologies, we showed that for certain reduction schemes, which we call the disk reduction schemes, we get $2d$ $BF$-theories that have close relations with the boundary CFTs in Chern-Simons theories. In section \ref{sec 5: Disk Reduction of Chern-Simons Theories on X = R * D2}, we constructed these disk reductions, where we studied Chern-Simons theories on the topology $X = \mathbb R \times D_2$, with $D_2$ a disk. We showed that by a disk reduction, one obtains a $BF$-theory on a 2-dimensional spacetime. It would be interesting to investigate the relation of this reduced theory to rational conformal field theories (RCFT), that appear on the boundary $\mathbb R \times S^1$ of $U(1)$ Chern-Simons theory through the edge modes. The reduced actions for gauge groups $U(1)$ and $SU(2)$ give $BF$-theories on the spacetime $I_R \times S^1$ (or $S^1_R \times S^1$) with gauge groups $U(1)$ and $U(1)\times U(1)$, respectively. These $BF$-actions can be put in the same form as the boundary term of the $U(1)$ Chern-Simons theory on $X$, which signals that the symmetry reductions in the bulk theory have a close relationship with CFTs, in addition to the boundary modes of Chern-Simons theory which famously give CFTs. Thus, it is natural to suspect that the symmetry reductions have a relation with holography. These results call for an involved investigation containing Chern-Simons theories, CFTs, 3d gravity theories, and probably Topological Field Theories (TFTs) defined over $X= \mathbb R \times \Sigma$ \cite{Kapustin:2010if, Witten:1988hf, Elitzur-Shmuel-Moore-Schwimmer-Seiberg:1989nr, Dijkgraaf:1989pz}. \\

The second part of the paper was concerned with Chern-Simons-Higgs theories and their symmetry reductions, which have been studied in the context of Chern-Simons-Higgs monopoles \cite{Edelstein:1994dy}. It is well known that in these theories monopoles with non-trivial flux must necessarily have divergent Euclidean action, as we demonstrated in section \ref{sec 6: SU(2) Chern-Simons Higgs and their monopoles}. Upon symmetry reduction by spherically symmetric ansatz for monopoles, we get a 1-dimensional action that governs these monopoles. Using the conformal equivalence $\mathbb R^3-\{0\} \sim \mathbb R^+ \times S^2$, we showed that Chern-Simons-Higgs theory defined over this space reduces, under certain circumstances, to a 1-dimensional theory which can be identified at the classical level with a supersymmetric quantum mechanics. We then showed that the reduced theory can be put in a form that contains fermions, at the level of the partition function. The action we obtain of which the partition function is equivalent to that of the symmetry reduction of Chern-Simons-Higgs theory is almost a supersymmetric action, although there is one term that breaks the supersymmetry. Similar conclusions hold for the $SL(2,\mathbb C)$ Chern-Simons Higgs theory, with two distinct actions that are copies of each other. We finally discuss the semiclassical regime of the Chern-Simons-Higgs theory. We argue that it is not likely to get an exact agreement at the semiclassical limit as we got for the pure Chern-Simons theory due to the presence of the Higgs field, although the reduced theory captures some portion of the original Chern-Simons-Higgs theory in the semiclassical limit. \\

With all these results, questions, and possible directions to follow that were unearthed from the study of symmetry reductions in the context of Chern-Simons gauge theories, we remark that these ideas should be investigated under more scrutiny, in particular, to understand the relationships between the quantum theories of the reduced actions and those of the original ones. One may use these results to study the saddle points of the gravitational path integrals and understand the relation between the 3-dimensional gravity and boundary CFTs, which was recently investigated in light of the correspondence with double-scaled SYK models \cite{Verlinde:2024zrh, Verlinde:2024znh, Narovlansky:2023lfz, Susskind:2022bia}. Our reduction scheme may be applied to these investigations. It would also be interesting to integrate the concept of symmetry reductions into the framework of generalized symmetries \cite{Gaiotto:2014kfa, Shao:2023gho, Bhardwaj:2023kri, Schafer-Nameki:2023jdn, Gomes:2023ahz, Brennan:2023mmt}, which is under extensive development. Most notably, understanding how $1-$form symmetries behave under symmetry reductions would allow a clear investigation of how the observables of the Chern-Simons theory are related to the observables of the reduced theory. \\

\section*{Acknowledgements} 
We would like to thank Batuhan Kaynak Acar, Çağdaş Ulus Ağca, Altay Etkin, Arda Hasar, Seçkin Kürkçüoğlu, and Farbod-Sayyed Rassouli for discussions and raising helpful questions about our work. We thank Prof. Dr. Sakir Erkoc for support.

\appendix 

\section{Spherically Symmetric ans\"atze for Non-Abelian Chern-Simons Theory and the Dreibein }

Throughout the paper, we employ the following ans\"atze:
\begin{equation}
    A_i^a =   \varepsilon_{i}^{\hspace{2mm} ak} \frac{x_k}{r^2} \big(\varphi_2(r) - 1 \big) + \frac{\delta^{\perp a}_{i}}{r} \varphi_1(r) + \frac{x_ix^a}{r^2} A(r),
\end{equation}
\begin{equation}
    e_i^a = -\varepsilon^a_{\hspace{2mm}ik} \frac{x^k}{r^2} e_1(r) + \delta^{\perp a}_i \frac{e_2(r)}{r} + \frac{x_ix^a}{r^2} e_3(r),
\end{equation}
\begin{equation}
    \Phi^a = \frac{x^a}{r} \Phi(r),
\end{equation}
where $ \delta^{\perp a}_i = \left( \delta_i^a - \frac{x_ix^a}{r^2} \right)$. We note the following identities that will be helpful
\begin{equation}  \label{Identities among varepsilon, deltaperp, and xi xa term}
\begin{aligned}
    \varepsilon_{iak} \delta^{\perp ak} = 0 &= \varepsilon_{iak} x^a x^k, \\
    \delta_{i}^{\perp a} x_i = 0  \quad ; \quad 
    \delta_i^{\perp a} \delta_a^{\perp j} &= \delta_i^{\perp j}  \quad ; \quad \delta_i^{\perp i} = 2, \\
    \varepsilon_{ajl} \delta_i^{\perp a} x^l &= \varepsilon_{ijl} x^l,
    \end{aligned}
\end{equation}
along with standard rules of calculus such as $\partial_i f(r) = \frac{x_i}{r} f'$. We will mean a derivative with respect to $r$ whenever there is a prime. If there is a dot, it is a derivative with respect to either time $t$ or the Euclidean time $\tau$. With these, one calculates
\begin{equation}
    \varepsilon^{ijk} \partial_j A_k^a = \varepsilon^{iak} \frac{x_k}{r^2} \Big(- (\varphi'_1 - A) \Big)  +  \frac{\delta^\perp_{ia}}{r} \Big( \varphi'_2 \Big) + \frac{x_ix_a}{r^4} \Big( 2\left( \varphi_2-1  \right) \Big),
\end{equation}
\begin{equation}
    \varepsilon^{ijk} \varepsilon_{abc} A_j^b A_k^c = \varepsilon^{i}_{\hspace{2mm} ak} \frac{x^k}{r^2} \Big( 2(\varphi_2 - 1) A \Big) +  \frac{\delta^{\perp i}_{a}}{r} \Big( 2\varphi_1 A \Big) + \frac{x^ix_a}{r^4} \Big( 2\left( (\varphi_2-1)^2 + \varphi_1^2  \right) \Big).
\end{equation}
Contracting these with $A_i^a$, one gets
\begin{equation}
\begin{aligned}
    \varepsilon^{ijk} A_i^a \partial_j A_k^a &= \frac{2}{r^2} \Big( -(\varphi_2 - 1) (\varphi'_1 - A) + \varphi_1 \varphi'_2 + A (\varphi_2-1) \Big) \\
    &= \frac{2}{r^2} \Big( \varphi_1 \varphi'_2 - \varphi'_1 (\varphi_2-1) + 2A (\varphi_2-1) \Big),
\end{aligned}
\end{equation} 
\begin{equation}
\begin{aligned}
    \varepsilon^{ijk} \varepsilon_{abc} A_i^a A_j^b A_k^c &= \frac{2}{r^2} \Big( 2A (\varphi_2-1)^2 + 2A \varphi_1^2 + A \big( (\varphi_2-1)^2 + \varphi_1^2 \big) \Big) \\
    &= \frac{6}{r^2} A \Big( (\varphi_2-1)^2 + \varphi_1^2 \Big).
\end{aligned}
\end{equation} 
With this one can find the field strength tensor $F_{ij}^a = \partial_i A_j^a - \partial_j A_i^a + \varepsilon^{abc} A_i^b A_j^c$:
\begin{equation}
    \frac{1}{2} \varepsilon^{ijk} F_{jk}^a = -\left( \frac{\partial_1\varphi_1  - A \varphi_2}{r^2} \right) \varepsilon_{iak} x_k + \left( 
    \frac{\partial_1\varphi_2 + A \varphi_1}{r} \right) \delta^{\perp}_{ia} - \left( \frac{1 - \varphi_1^2 - \varphi_2^2}{r^4} \right) x_ix_a,
\end{equation}
Moreover, the Chern-Simons form reads
\begin{equation} \label{SU(2) CS form after the spherical ansatz}
\begin{aligned}
    \varepsilon^{ijk} \left( A_i^a\partial_j A_k^a + \frac{1}{3} \varepsilon^{abc} A_i^a A_j^b A_k^c \right) &= \frac{2}{r^2} \Big( \varphi_1 \varphi'_2 - \varphi'_1 (\varphi_2-1) + A \big( (\varphi_2-1)^2 + 2(\varphi_2-1) + \varphi_1^2 \big)\Big) \\
    &= \frac{2}{r^2} \bigg( \varphi_1 \varphi'_2 - \varphi'_1 (\varphi_2-1) + A (\varphi_2^2 + \varphi_1^2 - 1) \bigg).
\end{aligned}
\end{equation}
For the dreibein, one has
\begin{equation}
    \varepsilon^{ijk} \varepsilon_{abc} e_j^b e_k^c =  \varepsilon^{iak} \frac{x_k}{r^2} \Big( 2 e_1 e_3 \Big) +  \frac{\delta^{\perp ia} }{r} \Big( 2e_2 e_3 \Big) + \frac{x^i x^a}{r^4} \Big( 2 \left( e_1^2 + e_2^2 \right) \Big),
\end{equation}
contracting this with $e_i^a$, one gets
\begin{equation}
    e \equiv \det e = \frac{1}{3!} \varepsilon^{ijk} \varepsilon_{abc} e_i^a e_j^b e_k^c = \frac{1}{r^2} e_3 \Big( e_1^2 + e_2^2 \Big).
\end{equation}
One can find the metric using
\begin{equation}
\begin{aligned}
    g_{ij} &= e_i^a e_j^b \delta_{ab} = e_i^a e_{ja} \\
    & =\left( -\varepsilon^a_{\hspace{2mm}ik} \frac{x^k}{r^2} e_1(r) + \delta^{\perp a}_i \frac{e_2(r)}{r} + \frac{x_ix^a}{r^2} e(r) \right) \left( -\varepsilon_{ajl} \frac{x^l}{r^2} e_1(r) + \delta^\perp_{ja} \frac{e_2(r)}{r} + \frac{x_jx_a}{r^2} e(r) \right).
\end{aligned}
\end{equation}
Explicitly, this reads
\begin{equation}
\begin{aligned}
     g_{ij} &= \varepsilon^a_{\hspace{2mm}ik} \varepsilon_{ajl} \frac{x^k x^l}{r^4} e_1^2 -\varepsilon^a_{\hspace{2mm}ik} \delta^{\perp}_{ja} \frac{x^k}{r^2} e_1(r) e_2(r) -\varepsilon^a_{\hspace{2mm}ik} \frac{x^kx_jx_a}{r^4} e_1(r) e_3(r) \\
     & - \delta^{\perp a}_i \varepsilon_{ajl} \frac{x^l}{r^3} e_1(r) e_2(r) + \delta^{\perp a}_i \delta^\perp_{ja}  \frac{e^2_2(r)}{r^2} + \delta^{\perp a}_i \frac{x_jx_a}{r^2} e_2(r) e_3(r) \\
     & - \varepsilon_{ajl} \frac{x_ix^ax^l}{r^4} e_3(r)e_1(r) + \delta^\perp_{ja} \frac{x_ix^a}{r^3} e_3(r) e_2(r) + \frac{x_ix_j x_a x^a}{r^4} e_3^2(r).
\end{aligned}
\end{equation}
Using the identities in (\ref{Identities among varepsilon, deltaperp, and xi xa term}), one arrives at
\begin{equation}
\begin{aligned}
    g_{ij} &= (\delta_{ij} \delta_{kl} - \delta_{il} \delta_{kj}) \frac{x^kx^l}{r^4} e_1^2(r) - \varepsilon_{jil} \frac{x^l}{r^2} e_1(r) e_2(r) - \varepsilon_{ijl} \frac{x^l}{r^2} e_1(r) e_2(r) + \delta^\perp_{ij} e^2_2(r) + \frac{x_i x_j}{r^2} e_3^2 \\
    &= \frac{1}{r^2} \left( \delta^\perp_{ij} (e_1^2 + e_2^2) + x_ix_j e_3^2 \right) \\
    &= \frac{e_1^2+e_2^2}{r^2} \delta_{ij} + \left( e_3^2 - \frac{e_1^2+e_2^2}{r^2} \right) \frac{x_ix_j}{r^2}.
\end{aligned}
\end{equation}
One can also compute the inverse dreibein, which is defined by the consistency condition
\begin{equation}
    \delta_i^j = e_i^a E^j_a.
\end{equation}
Let us write $E$ as:
\begin{equation}
    E_a^j = - \varepsilon_a^{\hspace{2mm} jk} \frac{x_k}{r^2} E_1 + \delta^{\perp j}_a \frac{E_2}{r} + \frac{x^jx_a}{r^2} E_3,
\end{equation}
thus, one has
\begin{equation}
    \delta_i^j = e_i^a E^j_a = \left( -\varepsilon^a_{\hspace{2mm}ik} \frac{x^k}{r^2} e_1 + \delta^{\perp a}_i \frac{e_2}{r} + \frac{x_ix^a}{r^2} e_3 \right) \left( - \varepsilon_a^{\hspace{2mm} jl} \frac{x_l}{r^2} E_1 + \delta^{\perp j}_a \frac{E_2}{r} + \frac{x^jx_a}{r^2} E_3 \right),
\end{equation}
explicitly, this is
\begin{equation}
\begin{aligned}
    \delta_i^j &= \varepsilon^a_{\hspace{2mm}ik} \varepsilon_a^{\hspace{2mm} jl} \frac{x^kx_l}{r^4} e_1E_1 - \varepsilon^a_{\hspace{2mm}ik} \delta^{\perp j}_a \frac{x^k}{r^3} e_1E_2 - \varepsilon^{a}_{\hspace{2mm}ik} \frac{x^kx^jx_a}{r^4}e_1 E_3 \\
    &  - \varepsilon_a^{\hspace{2mm} jk} \delta^{\perp a}_i \frac{x_k}{r^3} e_2E_1 \hspace{3mm} + \delta^{\perp j}_a \delta^{\perp a}_i \frac{e_2E_2}{r^2} \hspace{5mm} + \delta_i^{\perp a} \frac{x^jx_a}{r^3} e_2 E_3 \\
    & -\varepsilon_a^{\hspace{2mm}jk} \frac{x_kx_ix^a}{r^4} e_3 E_1 \hspace{2mm} + \delta^{\perp j}_a \frac{x_ix^a}{r^3} e_3 E_2 \hspace{4mm} + \frac{x_i x^a x^j x_a}{r^4} e_3 E_3.
\end{aligned}
\end{equation}
Using the identities in (\ref{Identities among varepsilon, deltaperp, and xi xa term}), one gets
\begin{equation}
\begin{aligned}
    \delta_i^j &= (\delta_i^j \delta^l_k - \delta^l_i \delta^j_k) \frac{x^k x_l}{r^4} e_1E_1 - \varepsilon^j_{\hspace{2mm} ik} \frac{x^k}{r^3} e_1E_2 - 0 \\
    &  - \varepsilon_i^{\hspace{2mm} jk} \frac{x_k}{r^3} e_2E_1  + \delta^{\perp j}_i \frac{e_2E_2}{r^2}  + 0\\
    & + 0 + 0 + \frac{x_i x^j}{r^2} e_3 E_3.
\end{aligned}
\end{equation}
Simplifying further
\begin{equation}
\begin{aligned}
    \delta^j_i &= \delta_i^j \frac{e_1E_1}{r^2} - \frac{x^jx_i}{r^4} e_1E_1 - \varepsilon^j_{\hspace{2mm}ik} \frac{x^k}{r^3} (e_1E_2 - e_2 E_1) + \delta_i^j \frac{e_2E_2}{r^2} - \frac{x_ix^j}{r^4} e_2E_2 + \frac{x_ix^j}{r^2} e_3 E_3 \\
    &= \delta^j_i \frac{e_1E_1 + e_2E_2}{r^2} + \frac{x_ix^j}{r^2} \left( e_3 E_3 - \frac{e_1E_1+e_2E_2}{r^2} \right) - \varepsilon^j_{\hspace{2mm}ik} \frac{x^k}{r^3} (e_1E_2 - e_2 E_1).
\end{aligned}
\end{equation}
This gives three equations
\begin{equation} 
\begin{aligned}
    r^2 &= e_1E_1 + e_2E_2 , \\
    0 &= e_3 E_3 - \frac{e_1E_1+e_2E_2}{r^2}, \\
    0 &= e_1E_2 - e_2 E_1, \\
\end{aligned}
\end{equation}
Using the first one in the second gives
\begin{equation}
    E_3 = \frac{1}{e_3}.
\end{equation}
Using the third equation, we write $E_2 = \frac{E_1e_2}{e_1}$ and input this into the first one 
\begin{equation}
    r^2 = e_1 E_1 + \frac{e_2^2E_1}{e_1} =
    \frac{(e_1^2+e_2^2)E_1}{e_1},
\end{equation}
hence
\begin{equation}
    E_1 = e_1 \frac{r^2}{e_1^2+e_2^2} \quad ; \quad E_2 = e_2 \frac{r^2}{e_1^2+e_2^2}.
\end{equation}
So, all three ansatz functions that constitute a spherically symmetric inverse of the dreibein $E_a^i$ are found. The form of $E_a^i$ in terms of little $e$'s is given by
\begin{equation}
    E_a^j = - \varepsilon_a^{\hspace{2mm} jk} x_k \frac{e_1}{e_1^2+e_2^2} + \delta^{\perp j}_a \frac{r e_2}{e_1^2+e_2^2} + \frac{x^jx_a}{r^2} \frac{1}{e_3}.
\end{equation}
It is almost a trivial manner to compute the inverse of the metric now. We will use the structural similarity of $e$ and $E$ to write
\begin{equation}
    g^{ij} = E^i_a E^j_b \delta^{ab} = \frac{1}{r^2} \left( \delta^{\perp ij} (E_1^2+E_2^2) + x^i x^j E_3^2 \right),
\end{equation}
in terms of little $e$'s
\begin{equation}
    g^{ij} = \delta^{ij} \frac{r^2}{e_1^2+e_2^2} + x^i x^j \left( \frac{1}{r^2 e_3^2} - \frac{1}{e_1^2 + e_2^2} \right).
\end{equation}
We note that
\begin{equation}
    x^a \equiv e_i^a x^i.
\end{equation}
On the other hand, if we use the spherically symmetric ansatz, we get
\begin{equation}
    x^a = e_3 x^a,
\end{equation}
which compels us to set $e_3 = 1$. The metric now reads
\begin{equation}
    g_{ij} = \frac{e_1^2+e_2^2}{r^2} \delta_{ij} + \left( 1 - \frac{e_1^2+e_2^2}{r^2} \right) \frac{x_ix_j}{r^2},
\end{equation}
and the line element
\begin{equation}
\begin{aligned}
    ds^2 &= \frac{e_1^2+e_2^2}{r^2} (dr^2 + r^2 d\Omega^2) + \left( 1 - \frac{e_1^2+e_2^2}{r^2} \right) dr^2 \\
    &= dr^2 + (e_1^2+e_2^2) d\Omega^2.
\end{aligned}
\end{equation}
Where we've used $dr = x_i dx^i/r$.\\

Let us now record some results involving the adjoint scalar field
\begin{equation}
    \partial_i \Phi^a = \delta^{\perp a}_{i} \frac{\Phi}{r} + \frac{x_ix^a}{r^2} \Phi',
\end{equation}
\begin{equation}
\begin{aligned}
    \varepsilon^{abc} A_i^b \Phi^c &= \varepsilon^{abc} \left( (\varphi_2 - 1) \varepsilon_{i}^{\hspace{2mm} bk} \frac{x_k}{r^2}  + \varphi_1 \frac{\delta^{\perp b}_{i}}{r} +  A \frac{x_ix^b}{r^2} \right) \frac{x^c}{r} \Phi \\
    &= ( \delta^{a}_{i} \delta_{ck} - \delta^{a}_{k} \delta_{ci} ) \frac{x_kx_c}{r^3} (\varphi_2-1) \Phi - \varepsilon_{i}^{\hspace{2mm} ac} \frac{x_c}{r^2} \varphi_1 \Phi + \varepsilon^{abc} \frac{x_ix_bx_c}{r^2} \left(  \cdots \right) \\
    &= \frac{\delta^{\perp a}_{i} }{r} (\varphi_2-1)\Phi - \varepsilon_{i}^{\hspace{2mm} ac} \frac{x_c}{r^2} \varphi_1 \Phi.
\end{aligned}
\end{equation}
Thus, the covariant derivative is given by
\begin{equation}
\begin{aligned}
    D_i \Phi^a &= \partial_i \Phi^a + \varepsilon^{abc} A_i^b \Phi^c \\ 
    &= \varepsilon_{i}^{\hspace{2mm} ac} \frac{x_c}{r^2} \Big( - \varphi_1 \Phi \Big) + \frac{\delta^{\perp a}_{i}}{r} \Big( \varphi_2 \Phi \Big) + \frac{x_ix^a}{r^2} \Big( \Phi' \Big).
\end{aligned}
\end{equation} 
One computes the following contractions
\begin{equation}
\begin{aligned}
    \frac{1}{2} \delta^{ij} D_i \Phi^a D_j \Phi^a &= \frac{1}{r^2} \left( \frac{r^2}{2} \Phi'^2 + (\varphi_1^2 + \varphi_2^2) \Phi^2 \right), \\
    \frac{1}{2} x^i x^j D_i \Phi^a D_j \Phi^a &= \frac{1}{2} \Phi'^2 .
\end{aligned}
\end{equation}
Therefore, the kinetic term of a Higgs field in a spherically symmetric space reduces to
\begin{equation}
\begin{aligned}
    \frac{1}{2} g^{ij} D_i \Phi^a D_j \Phi^a &= \frac{1}{2} \frac{r^2}{e_1^2+e_2^2} \delta^{ij} D_i \Phi^a D_j \Phi^a + \frac{1}{2} \left( \frac{1}{r^2} - \frac{1}{e_1^2+e_2^2} \right) x^i x^j D_i \Phi^a D_j \Phi^a \\
    &= \frac{1}{2} \Phi'^2 + \frac{1}{e_1^2+e_2^2} (\varphi_1^2 + \varphi_2^2) \Phi^2.
\end{aligned}
\end{equation}
In particular, one has
\begin{equation}
\begin{aligned}
    \int d^3x \sqrt{|g|} \frac{1}{2} g^{ij} D_i \Phi^a D_j \Phi^a &= 4\pi \int dr (e_1^2+e_2^2) \left(  \frac{1}{2} \Phi'^2 + \frac{1}{e_1^2+e_2^2} (\varphi_1^2 + \varphi_2^2) \Phi^2 \right) \\
    &= 4\pi \int dr \left( \frac{1}{2} (e_1^2+e_2^2) \Phi'^2 + (\varphi_1^2+ \varphi_2^2) \Phi^2 \right).
\end{aligned}
\end{equation}  
Similarly, for any spherically symmetric integrand, it follows that
\begin{equation}
    \int d^3x \sqrt{|g|} \Big[ \cdots \Big] = 4\pi \int dr r^2 \frac{e_1^2+e_2^2}{r^2} \Big[\cdots \Big] = 4\pi \int dr (e_1^2+e_2^2) \Big[ \cdots \Big].
\end{equation}
In computing the abelian field strength in the direction of symmetry breaking, we will also need
\begin{equation}
\begin{aligned}
    \mathcal F_{ij}^{(2)} &= - \frac{1}{\Phi^3} \varepsilon^{abc} \Phi^a D_i \Phi^b D_j \Phi^c \\
    &= -\frac{1}{\Phi^3} \varepsilon^{abc} \Phi \frac{x^a}{r} 
    \left( \varepsilon_{i}^{\hspace{2mm} bd} \frac{x_d}{r^2} \Big( - \varphi_1 \Phi \Big) + \frac{\delta^{\perp b}_{i}}{r} \Big( \varphi_2 \Phi \Big) + \frac{x_ix^b}{r^2} \Big( \Phi' \Big) \right) \\
    & \hspace{23mm} \times \left( \varepsilon_{j}^{\hspace{2mm} ce} \frac{x_e}{r^2} \Big( - \varphi_1 \Phi \Big) + \frac{\delta^{\perp c}_{j} }{r} \Big( \varphi_2 \Phi \Big) + \frac{x_jx^c}{r^2} \Big( \Phi' \Big) \right).
\end{aligned}
\end{equation}
We will first compute the corresponding magnetic field
\begin{equation}
    \mathcal B_k^{(2)} = \frac{1}{2\Phi^3} \varepsilon_{jik} \mathcal F_{ij}^{(2)} = -\frac{1}{2 \Phi^3} \varepsilon_{kij} \varepsilon^{abc} \Phi^a D_i \Phi^b D_j \Phi^c.
\end{equation}
One finds
\begin{equation}
    \frac{1}{2} \varepsilon_{kij} \varepsilon^{abc} D_i \Phi^b D_j \Phi^c = \varepsilon_k^{\hspace{2mm} ad} \frac{x_d}{r^2} \Big( \varphi_1 \Phi \Phi' \Big) + \frac{\delta_k^{\perp a}}{r} \Big( \varphi_2 \Phi \Phi' \Big) + \frac{x_k x^a}{r^4} \Big( (\varphi_1^2 + \varphi_2^2) \Phi^2 \Big), 
\end{equation}
\begin{equation}
    \Longrightarrow \frac{1}{2} \varepsilon_{kij} \varepsilon^{abc} \Phi^a D_i \Phi^b D_j \Phi^c = \frac{x_k}{r^3} \Phi^3 (\varphi_1^2+\varphi_2^2),
\end{equation} 
so the magnetic field reads
\begin{equation}
    \mathcal B_k^{(2)} = -\frac{x_k}{r^3} (\varphi_1^2 + \varphi_2^2).
\end{equation}
Now, observe that
\begin{equation}
\begin{aligned}
    \varepsilon_{klm} \mathcal B_k^{(2)} = \frac{1}{2}  \varepsilon_{klm} \varepsilon_{kij} \mathcal F_{ij}^{(2)} &= \frac{1}{2} (\delta_{il} \delta_{jm} - \delta_{im} \delta_{jl}) \mathcal F_{ij}^{(2)} \\
    &= \mathcal F_{lm}^{(2)} ,
\end{aligned}
\end{equation}
therefore, the abelian field strength is given by
\begin{equation}
    \mathcal F_{ij}^{(2)} = \varepsilon_{ijk} \mathcal B_k^{(2)} = -\varepsilon_{ijk} \frac{x_k}{r^3} (\varphi_1^2+\varphi_2^2).
\end{equation}

\section{Disk Reduction Ansatz for \texorpdfstring{$U(1)$}{U(1)} Chern-Simons Theory}
We take a $U(1)$ Chern-Simons theory on the disk topology $X = \mathbb R \times D_2$ and make the following reduction
\begin{equation}
\begin{aligned}
    A_i &= \varepsilon_{ij} \frac{x_j}{r^2} \phi(r,t) + \frac{x_i}{r} A_1(r,t), \\
    A_t &= A_0(r,t),
\end{aligned}
\end{equation}
$(i = 1,2)$ and $r = (x_ix_i)^{1/2}$. With this, one has
\begin{equation}
\begin{aligned}
    \partial_j A_k &= \varepsilon_{kl} \partial_j \left( \frac{x_l}{r^2} \phi \right) + \partial_j \left( \frac{x_k}{r} A_1 \right) \\ 
    &= \varepsilon_{kl} \left( \frac{\delta_{jl}}{r^2} - 2 \frac{x_jx_l}{r^4} \right) \phi + \varepsilon_{kl} \frac{x_jx_l}{r^3} \phi' + \left( \frac{\delta_{jk}}{r} - \frac{x_jx_k}{r^3} \right) A_1 + \frac{x_jx_k}{r^2} A_1' \\
    &= \left( \frac{\varepsilon_{kj}}{r^2} - 2\varepsilon_{kl} \frac{ x_jx_l}{r^4} \right) \phi + \varepsilon_{kl} \frac{x_jx_l}{r^3} \phi'  +  \left( \frac{\delta_{jk}}{r} - \frac{x_jx_k}{r^3} \right) A_1 + \frac{x_jx_k}{r^2} A_1'. 
\end{aligned}
\end{equation}
Then
\begin{equation}
\begin{aligned}
    A_t \varepsilon^{0jk} \partial_j A_k &= \left( \varepsilon_{0jk} \varepsilon_{kj}\frac{1}{r^2} - 2  \varepsilon_{0jk} \varepsilon_{kl} \frac{x_jx_l}{r^4} \right) A_0 \phi + \varepsilon_{0jk} \varepsilon_{kl} \frac{x_jx_l}{r^3} A_0 \phi' \\
    &= \left( -\frac{2}{r^2} + \frac{2}{r^2} \right) A_0 \phi + \varepsilon_{0jk} \varepsilon_{kl} \frac{x_jx_l}{r^3} A_0 \phi' \\
    &= -\frac{A_0 \phi'}{r},
\end{aligned}
\end{equation}
and
\begin{equation}
\begin{aligned}
    -\varepsilon^{0ij} A_i \partial_0 A_j &= -\varepsilon^{0ij} \left( \varepsilon_{il} \frac{x_l}{r^2} \phi + \frac{x_i}{r} A_1 \right) \left( \varepsilon_{jm} \frac{x_m}{r^2} \dot\phi + \frac{x_j}{r} \dot A_1 \right) \\
    &= - \left( \frac{x_j}{r^2} \phi + \varepsilon_{ij} \frac{x_i}{r} A_1 \right) \left( \varepsilon_{jm} \frac{x_m}{r^2} \dot\phi + \frac{x_j}{r} \dot A_1 \right)\\
    &= - \frac{\phi \dot A_1}{r} + \frac{A_1 \dot\phi}{r}
\end{aligned}
\end{equation}
and finally
\begin{equation}
\begin{aligned}
    \varepsilon^{ij0} A_i \partial_j A_t &= \varepsilon^{ij0} \left( \varepsilon_{il} \frac{x_l}{r^2} \phi + \frac{x_i}{r} A_1 \right) \frac{x_j}{r} A_0'  \\
    &= \left( \frac{x_j}{r^2} \phi + \varepsilon^{ij} \frac{x_i}{r} A_1  \right) \frac{x_j}{r} A'_0 \\
    &= \frac{\phi A'_0}{r}.
\end{aligned}
\end{equation}
Hence, the Chern-Simons form reduces to
\begin{equation}
    CS(A) = \varepsilon^{\mu\nu\rho} A_\mu \partial_\nu A_\rho = \frac{1}{r} ( \phi F_{10} + A_1 \partial_0 \phi - A_0 \partial_1 \phi),
\end{equation}
with $F_{10} = \partial_r A_0 - \partial_0 A_1$. 

\section{Disk Reduction Ansatz for \texorpdfstring{$SU(2)$}{SU(2)} Chern-Simons Theory}
We consider $SU(2)$ Chern-Simons theory on $X= \mathbb R \times D_2$ with $D_2$ a disk of radius $R$. We make the following ansatz on the non-Abelian gauge field 
\begin{equation}
\begin{aligned}
    A^b_i &= \varepsilon_{ib}\ \frac{1}{r}\ \phi_A(r,t) + \frac{\left( \delta_{ib} - 2 \frac{x_ix_b}{r^2} \right)}{r} \chi_A(r,t) + \frac{x_ix_b}{r^2} A_r(r,t) , \\
    A^3_i &= \varepsilon_{ij} \frac{x_j}{r^2} \phi_B(r,t) + \frac{x_i}{r} B_r(r,t),
\end{aligned}
\end{equation}
($i$ and $b$ runs from $1$ to $2$). Since we are on the disk $D_2$ of radius $R$, $0<r\leq R$, using the ansatz above we get a reduced theory on $\mathbb R \times I_R$ with $I_R = (0,R)$. We consider the Chern-Simons action in the $A_0 = 0$ gauge
\begin{equation}
    S = \frac{\kappa}{8\pi} \int dt \int_{D_2} d^2x \varepsilon^{ij}\ A_i^a \frac{d}{dt} A_j^a.
\end{equation}
We have
\begin{equation}
\begin{aligned}
    \varepsilon^{ij} A_i^b \frac{d}{dt} A_j^b &= \varepsilon^{ij} 
    \left[ \varepsilon_{ib}\ \frac{1}{r}\ \phi_A(r,t) + \frac{\left( \delta_{ib} - 2 \frac{x_ix_b}{r^2} \right)}{r} \chi_A(r,t) + \frac{x_ix_b}{r^2} A_r(r,t) \right] \\
    & \hspace{9.4mm} \left[ \varepsilon_{jb}\ \frac{1}{r}\ \partial_t\phi_A(r,t) + \frac{\left( \delta_{jb} - 2 \frac{x_jx_b}{r^2} \right)}{r} \partial_t\chi_A(r,t) + \frac{x_jx_b}{r^2} \partial_t A_r(r,t) \right] \\
    &= 0 + 0 + \frac{1}{r} \phi_A \partial_t A_r \\
    & - 0 + 0 + 0 \\
    & - \frac{1}{r} A_r \partial_t \phi_A  + 0 + 0\\
    &= \frac{1}{r} ( 2\phi_A \partial_t A_r - \partial_t [\phi_A A_r] ),
\end{aligned}
\end{equation}
and
\begin{equation}
\begin{aligned}
    \varepsilon^{ij} A_i^3 \frac{d}{dt} A_j^3 &= \varepsilon^{ij} \left[ \varepsilon_{ik} \frac{x_k}{r^2} \phi_B(r,t) + \frac{x_i}{r} B_r(r,t) \right]  \left[ \varepsilon_{jl} \frac{x_l}{r^2} \partial_t\phi_B(r,t) + \frac{x_i}{r} \partial_t B_r(r,t) \right] \\
    &= 0 + \frac{1}{r} \phi_B \partial_t B_r - \frac{1}{r} B_r \partial_t \phi_B + 0 \\
    &= \frac{1}{r} ( 2\phi_B \partial_t B_r - \partial_t[\phi_B B_r]).
\end{aligned} 
\end{equation}
Thus, the reduced Chern-Simons action reads
\begin{equation}
    S = \frac{\kappa}{4\pi} \left[ \int_{\mathbb R \times I_R} d^2x\ \left( \phi_A \partial_t A_r + \phi_B \partial_t B_r \right) - 2 \int_{I_R} dr \left( \phi_A A_r + \phi_B B_r \right) \big|_{t=-\infty}^{t=\infty} \right].
\end{equation}

% Bibliography

%% [A] Recommended: using JHEP.bst file
%% \bibliographystyle{JHEP}
%% \b^{i\Lambda} \psiibliography{biblio.bib}

%% or
%% [B] Manual formatting (see below)
%% (i) We suggest to always provide author, title and journal data or doi:
%% in short all the informations that clearly identify a document.
%% (ii) please avoid comments such as "For a review'', "For some examples",
%% "and references therein" or move them in the text. In general, please leave only references in the bibliography and move all
%% accessory text in footnotes.
%% (iii) Also, please have only one work for each \bibitem.

\bibliographystyle {unsrt}
\bibliography{biblio.bib}

\begin{thebibliography}{10}

\bibitem{Palais:1979rca}
Richard~S. Palais.
\newblock {The principle of symmetric criticality}.
\newblock {\em Commun. Math. Phys.}, 69(1):19--30, 1979.

\bibitem{Witten:1976ck}
Edward Witten.
\newblock {Some Exact Multi - Instanton Solutions of Classical Yang-Mills Theory}.
\newblock {\em Phys. Rev. Lett.}, 38:121--124, 1977.

\bibitem{Manton:2004tk}
N.~S. Manton and P.~Sutcliffe.
\newblock {\em {Topological solitons}}.
\newblock Cambridge Monographs on Mathematical Physics. Cambridge University Press, 2004.

\bibitem{Tekin:2000fp}
Bayram Tekin.
\newblock {Multi - instantons in $R^4$ and minimal surfaces in $R^{(2,1)}$}.
\newblock {\em JHEP}, 08:049, 2000.

\bibitem{Comtet:1978ue}
A.~Comtet.
\newblock {Instantons and Minimal Surfaces}.
\newblock {\em Phys. Rev. D}, 18:3890, 1978.

\bibitem{Witten:1988hf}
Edward Witten.
\newblock {Quantum Field Theory and the Jones Polynomial}.
\newblock {\em Commun. Math. Phys.}, 121:351--399, 1989.

\bibitem{Dunne:1998qy}
Gerald~V. Dunne.
\newblock {Aspects of Chern-Simons theory}.
\newblock In {\em {Les Houches Summer School in Theoretical Physics, Session 69: Topological Aspects of Low-dimensional Systems, hep-th/9902115}}.

\bibitem{Tong:2016kpv}
David Tong.
\newblock {Lectures on the Quantum Hall Effect, arXiv:1606.06687}.

\bibitem{Witten:1988hc}
Edward Witten.
\newblock {(2+1)-Dimensional Gravity as an Exactly Soluble System}.
\newblock {\em Nucl. Phys. B}, 311:46, 1988.

\bibitem{Dunne:1996yb}
Gerald~V. Dunne, Ki-Myeong Lee, and Chang-hai Lu.
\newblock {On the finite temperature Chern-Simons coefficient}.
\newblock {\em Phys. Rev. Lett.}, 78:3434--3437, 1997.

\bibitem{Verlinde:2024zrh}
Herman Verlinde and Mengyang Zhang.
\newblock {SYK Correlators from 2D Liouville-de Sitter Gravity, arXiv:2402.02584}.

\bibitem{Verlinde:2024znh}
Herman Verlinde.
\newblock {Double-scaled SYK, Chords and de Sitter Gravity, arXiv:2402.00635}.

\bibitem{Narovlansky:2023lfz}
Vladimir Narovlansky and Herman Verlinde.
\newblock {Double-scaled SYK and de Sitter Holography,arXiv:2310.16994}.

\bibitem{Susskind:2022bia}
Leonard Susskind.
\newblock {De Sitter Space, Double-Scaled SYK, and the Separation of Scales in the Semiclassical Limit, arXiv:2209.09999}.

\bibitem{Kapustin:2010if}
Anton Kapustin and Natalia Saulina.
\newblock {Surface operators in 3d Topological Field Theory and 2d Rational Conformal Field Theory, arXiv:1012.0911}.
\newblock pages 175--198.

\bibitem{Elitzur-Shmuel-Moore-Schwimmer-Seiberg:1989nr}
Shmuel Elitzur, Gregory~W. Moore, Adam Schwimmer, and Nathan Seiberg.
\newblock {Remarks on the Canonical Quantization of the Chern-Simons-Witten Theory}.
\newblock {\em Nucl. Phys. B}, 326:108--134, 1989.

\bibitem{Dijkgraaf:1989pz}
Robbert Dijkgraaf and Edward Witten.
\newblock {Topological Gauge Theories and Group Cohomology}.
\newblock {\em Commun. Math. Phys.}, 129:393, 1990.

\bibitem{Edelstein:1994dy}
J.~D. Edelstein and F.~A. Schaposnik.
\newblock {Monopoles in nonAbelian Chern-Simons Higgs theory}.
\newblock {\em Nucl. Phys. B}, 425:137--149, 1994.

\bibitem{Hori:2003ic}
K.~Hori, S.~Katz, A.~Klemm, R.~Pandharipande, R.~Thomas, C.~Vafa, R.~Vakil, and E.~Zaslow.
\newblock {\em {Mirror symmetry}}, volume~1 of {\em Clay mathematics monographs}.
\newblock AMS, Providence, USA, 2003.

\bibitem{Dunne:2000if}
Gerald~V. Dunne and Bayram Tekin.
\newblock {Calorons in Weyl gauge}.
\newblock {\em Phys. Rev. D}, 63:085004, 2001.

\bibitem{Coleman:1978ae}
Sidney~R. Coleman.
\newblock {The Uses of Instantons}.
\newblock {\em Subnucl. Ser.}, 15:805, 1979.

\bibitem{Shifman:1994ee}
Mikhail~A. Shifman, editor.
\newblock {\em {Instantons in gauge theories}}.
\newblock 1994.

\bibitem{Witten:1989ip}
Edward Witten.
\newblock {Quantization of {Chern-Simons} Gauge Theory With Complex Gauge Group}.
\newblock {\em Commun. Math. Phys.}, 137:29--66, 1991.

\bibitem{Tekin:1998wu}
Bayram Tekin.
\newblock {Dimensional reduction of three-dimensional gravity, hep-th/9812083}.

\bibitem{Tekin:1999gn}
Bayram Tekin.
\newblock {On the relevance of singular solutions in dS(3) and AdS(3) gravity, hep-th/9902090}.

\bibitem{Ferstl:2000qp}
Andrew Ferstl, Bayram Tekin, and Victor Weir.
\newblock {Gravitating instantons in three-dimensional anti-de Sitter space}.
\newblock {\em Phys. Rev. D}, 62:064003, 2000.

\bibitem{Blau:1993tv}
Matthias Blau and George Thompson.
\newblock {Derivation of the Verlinde formula from Chern-Simons theory and the G/G model}.
\newblock {\em Nucl. Phys. B}, 408:345--390, 1993.

\bibitem{Blau:2006gh}
Matthias Blau and George Thompson.
\newblock {Chern-Simons theory on S1-bundles: Abelianisation and q-deformed Yang-Mills theory}.
\newblock {\em JHEP}, 05:003, 2006.

\bibitem{Beasley:2005vf}
Chris Beasley and Edward Witten.
\newblock {Non-Abelian localization for Chern-Simons theory}.
\newblock {\em J. Diff. Geom.}, 70(2):183--323, 2005.

\bibitem{Tekin:1998qy}
Bayram Tekin, Kamran Saririan, and Yutaka Hosotani.
\newblock {Complex monopoles in the Georgi-Glashow-Chern-Simons model}.
\newblock {\em Nucl. Phys. B}, 539:720--738, 1999.

\bibitem{Pisarski:1986gr}
Robert~D. Pisarski.
\newblock {Magnetic Monopoles in Topologically Massive Gauge Theories}.
\newblock {\em Phys. Rev. D}, 34:3851, 1986.

\bibitem{Maldacena:1997re}
Juan~Martin Maldacena.
\newblock {The Large N limit of superconformal field theories and supergravity}.
\newblock {\em Adv. Theor. Math. Phys.}, 2:231--252, 1998.

\bibitem{Witten:1998qj}
Edward Witten.
\newblock {Anti-de Sitter space and holography}.
\newblock {\em Adv. Theor. Math. Phys.}, 2:253--291, 1998.

\bibitem{Gubser:1998bc}
S.~S. Gubser, Igor~R. Klebanov, and Alexander~M. Polyakov.
\newblock {Gauge theory correlators from noncritical string theory}.
\newblock {\em Phys. Lett. B}, 428:105--114, 1998.

\bibitem{Gaiotto:2014kfa}
Davide Gaiotto, Anton Kapustin, Nathan Seiberg, and Brian Willett.
\newblock {Generalized Global Symmetries}.
\newblock {\em JHEP}, 02:172, 2015.

\bibitem{Shao:2023gho}
Shu-Heng Shao.
\newblock {What's Done Cannot Be Undone: TASI Lectures on Non-Invertible Symmetries, arXiv:2308.00747}.

\bibitem{Bhardwaj:2023kri}
Lakshya Bhardwaj, Lea~E. Bottini, Ludovic Fraser-Taliente, Liam Gladden, Dewi S.~W. Gould, Arthur Platschorre, and Hannah Tillim.
\newblock {Lectures on generalized symmetries}.
\newblock {\em Phys. Rept.}, 1051:1--87, 2024.

\bibitem{Schafer-Nameki:2023jdn}
Sakura Schafer-Nameki.
\newblock {ICTP lectures on (non-)invertible generalized symmetries}.
\newblock {\em Phys. Rept.}, 1063:1--55, 2024.

\bibitem{Gomes:2023ahz}
Pedro R.~S. Gomes.
\newblock {An introduction to higher-form symmetries}.
\newblock {\em SciPost Phys. Lect. Notes}, 74:1, 2023.

\bibitem{Brennan:2023mmt}
T.~Daniel Brennan and Sungwoo Hong.
\newblock {Introduction to Generalized Global Symmetries in QFT and Particle Physics, arXiv:2306.00912}.

\end{thebibliography}

\end{document}